\documentclass{aa}
\usepackage[varg]{txfonts}
\usepackage[colorlinks=true,linkcolor=blue,citecolor=blue]{hyperref}
\usepackage{graphicx}
\usepackage{amsmath,amsfonts,amssymb,mathrsfs}
\usepackage{aas_macros}
\usepackage{scalerel,stackengine}
\usepackage{url}
\stackMath

\usepackage[dvipsnames]{xcolor}

\newcommand\reallywidehat[1]{%
\savestack{\tmpbox}{\stretchto{%
  \scaleto{%
    \scalerel*[\widthof{\ensuremath{#1}}]{\kern-.6pt\bigwedge\kern-.6pt}%
    {\rule[-\textheight/2]{1ex}{\textheight}}
  }{\textheight}%
}{0.5ex}}%
\stackon[1pt]{#1}{\tmpbox}%
}
\begin{document} 

 \title{Trade off between segment density and IWA for high-contrast imaging of exoplanets with a large segmented space mission}
 \titlerunning{Optimal segment density for high-contrast imaging of exoplanets}
\authorrunning{Leboulleux et al.}
  \author{Lucie Leboulleux\inst{1}, Raphaël Pourcelot\inst{2}, Laurent Pueyo\inst{3}, Bryony Nickson\inst{3}}
  \institute{Univ. Grenoble Alpes, CNRS, IPAG, 38000 Grenoble, France
  \and Max-Planck Institute for Astronomy (MPIA), Königstuhl 17, 69117 Heidelberg, Germany 
  \and Space Telescope Science Institute, 3700 San Martin Drive, Baltimore, MD 21218, USA
  \\
             \email{lucie.leboulleux@univ-grenoble-alpes.fr}} 

  \abstract
   {To image Earth-like exoplanets, the coronagraphs set up on future large space segmented telescopes such as the Habitable Worlds Observatory will need to access contrasts down to $10^{-10}$ at angular separations smaller than $100$ mas. This extreme requirement imposes an unprecedented constraint on segment phasing control, down to a few picometers.}
   {In this paper, we evaluate how this constraint can be released by optimizing several components of the telescope and instrument and quantifying their impact on the performance stability in the presence of segment phasing aberrations.}
   {We propose a system-level approach, with several parameters to adjust: the segmentation type of the primary mirror and the focal-plane mask size (or inner working angle). We compare the passive robustness to segment phasing errors of different systems, and the ability of a Zernike low-order wavefront sensor to reconstruct the aberrations and recover the target performance.}
   {Increasing the focal-plane mask radius or decreasing the number of segments across the pupil improves 1) the passive high-contrast performance robustness to segment-level errors: increasing the focal-plane mask radius from $3.5$ to $6.5\lambda/D$ relaxes phasing constraints by a factor of up to $4$ to maintain contrast near the IWA, and reducing the segment count from 85 (5 rings of segments) to 7 (1 ring of segments) relaxes them by a factor of up to $2$; and 2) the capability of the Zernike low-order wavefront sensor to reconstruct and correct for the errors, doubling its sensitivity to photon noise across the segment piston, tip and tilt modes (in our application case, increasing the focal-plane mask radius from $3.5$ to $6.5\lambda/D$ increases in average the segment mode coefficients by a factor $\sim 2$).}
   {The coronagraph focal-plane mask acts as a spatial high-pass filter whose cutoff frequency, set by its radius, determines which spatial frequencies of the segment phasing error power spectral density leak into the dark zone and which are detectable by the low-order wavefront sensor. Jointly optimizing the segmentation scheme and the focal-plane mask size, so that the low-order envelope of the power spectral density is blocked by the mask, can significantly relax segment phasing requirements, complementing active wavefront correction, with direct implications for the Habitable Worlds Observatory.}
    
   \keywords{Exoplanets - high-contrast imaging - coronagraphy - error budget - focal-plane wavefront sensing}

   \maketitle

\section{Introduction}
\label{s:Introduction}

The development of large segmented telescopes, particularly for high-contrast imaging applications, has made segment alignment and co-phasing errors a central concern, as achieving the targeted scientific goals is strongly linked to the wavefront quality. A representative example is the James Webb Space Telescope (JWST), for which a dedicated, month-long phasing procedure was required to reach a segment alignment accuracy of approximately 50 nm rms \citep{Acton2022, McElwain2023}. While this level of segment phasing is sufficient for JWST’s primary science objectives, it also highlights the increasing complexity of wavefront control in segmented apertures, particularly for space telescopes. 

Looking ahead, future flagship missions push these requirements dramatically further. In particular, the Habitable Worlds Observatory (HWO) aims to directly image Earth-like planets around Sun-like stars, targeting coronagraphic contrasts as low as $10^{-10}$ at angular separations below 100 mas. Meeting such extreme optical alignment levels implies controlling wavefront errors at the $10$ pm rms level \citep{Laginja2019b}, i.e., several orders of magnitude beyond current capabilities.

Achieving such extreme stability will not be possible without efficient wavefront sensing and control of segment-level aberrations. Among the various wavefront sensors (WFS), the Zernike Wavefront Sensor (ZWFS) \citep{Bloemhof2003, Wallace2011, N'Diaye2013} has demonstrated picometer-level sensitivity combined with excellent photon efficiency, making it a particularly promising candidate for HWO \citep{Chambouleyron2021, Ruane2020, Steeves2020}. In parallel, coronagraphic architectures employing a focal-plane mask (FPM), such as classical Lyot coronagraphs or Apodized Pupil Lyot Coronagraphs (APLCs), offer an additional opportunity for wavefront sensing: as the mask effectively acts as a high-pass spatial filter towards the coronagraphic arm, the fraction of the stellar light rejected by the FPM encodes information on low-order aberrations. Several WFS concepts, including ZWFS-based ones, have been optimized to exploit this previously unused signal for low-order wavefront sensing \citep{Pourcelot2022, Pourcelot2023}. 

This capability is especially relevant for segmented apertures, as segment-level piston, tip, and tilt errors predominantly manifest as low- or mid-order aberrations when the segments are sufficiently large \citep{Leboulleux2020}. For such segmented mirrors, the filtering by the FPM has two important consequences. First, the aberrated signal transmitted through the FPM strongly degrades the science performance: it impacts the Strehl ratio, and generates a photon floor at close angular separation, impacting the detectability of close-in planets; Second, the aberration signal reflected by the FPM must be efficiently collected by the low-order wavefront sensor to ensure a robust reconstruction of the segment-level aberrations.



These considerations raise a key system-level question: how should telescope segmentation and coronagraph design be jointly optimized to relax wavefront stability requirements and maximize the efficiency of the wavefront sensor? In this paper, we adopt an approach that explicitly accounts for the coupled telescope–instrument architecture. We investigate how two key parameters (the segment density, defined as the number of segments across the primary mirror, and the focal-plane mask diameter) affect both the coronagraph’s sensitivity to aberrations and the wavefront sensor’s ability to reconstruct them and therefore can be optimized to mitigate the impact of segment phasing errors on the contrast. Our analysis focuses on the APLC concept, one of the architectures capable of reaching the deepest contrasts \citep{Juanola-Parramon2022}. Further works could include the vector vortex coronagraph and the Phase-Induced Amplitude Apodization Complex Mask Coronagraph (PIAACMC) \citep{Belikov2018}.

Section \ref{s:Propagation of segment-level aberrations through the instrument} describes how segment-level aberration signals propagate through the optical system, both towards the coronagraphic arm (science camera) and the ZWFS. Sections \ref{s:Impact of the focal-plane mask radius and system cutoff frequency} and \ref{s:Impact of segment density} then examine how two key design parameters (the FPM radius and the segment density) affect the system’s robustness to aberrations. Finally, Section \ref{s:Conclusions} summarizes the main results and discusses the implications and limitations of this study.

\section{Propagation of segment-level aberrations through the instrument}
\label{s:Propagation of segment-level aberrations through the instrument}

\subsection{Localization of segment-level phasing errors and impact on the coronagraphic performance}

\begin{figure}
   \begin{center}
   \begin{tabular}{c}
   \includegraphics[width=8.5cm]{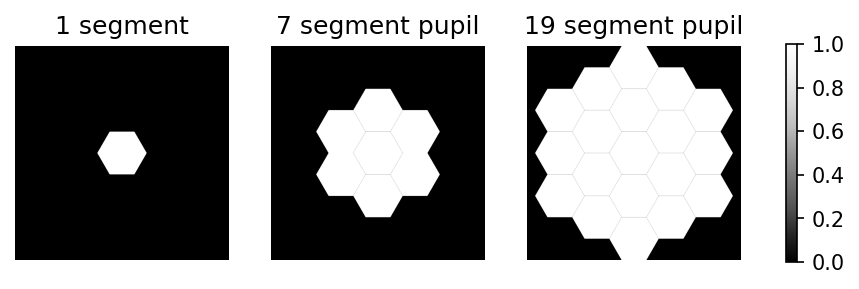} \\ 
   \includegraphics[width=8.5cm]{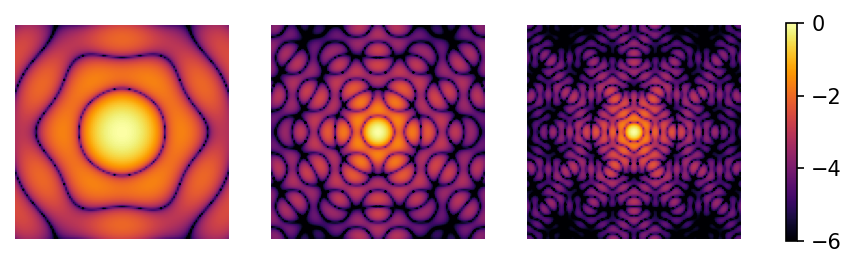}
   \end{tabular}
   \end{center}
   \caption[Fig] 
   { \label{fig:Figure0_PASTIS} Impact of the segment shape on the PSFs of segmented pupils: (left) Case of one single generic segment, whose PSF is also the low-order envelope of the PSF for any pupil composed of this segment; (center) Case of a pupil composed of 7 generic segments; (right) Case of a segmented pupil composed of 19 generic segments.}
   \end{figure}

In this paper, we consider a segmented pupil composed of a replication of one generic segment, as illustrated in Fig.~\ref{fig:Figure0_PASTIS} (top), combined with a coronagraph producing a high-contrast image~$I$. In the presence of segment-phasing errors described by a single Zernike polynomial (typically segment-level piston, tip, or tilt) and assuming monochromatic light at a wavelength~$\lambda$, this image can be expressed as follows \citep{Leboulleux2018, Laginja2019, Laginja2021, Laginja2022}:
\begin{equation}
\label{eq:PASTIS}
    I(\mathbf{u}) = \left \Vert \widehat{Z}(\mathbf{u}) \right \Vert ^2 \times \sum_{k_1=1}^{n_{\rm seg}} \sum_{k_2=1}^{n_{\rm seg}} c_{k_1,l} a_{k_1,l} c_{k_2,l} a_{k_2,l} \cos((\mathbf{r_{k_2}} - \mathbf{r_{k_1}} ). \mathbf{u})
\end{equation}
where $n_{\rm seg}$ is the total number of segments in the primary mirror, $\mathbf{u}$ is the position vector in the focal plane, $\widehat{Z}$ is the Fourier Transform of the segment-level Zernike polynomial $Z$ describing the error, $(c_k)_{k\in \lbrack 1,n_{\rm seg} \rbrack}$ are calibration coefficients accounting for the coronagraphic effect, $(a_k)_{k\in \lbrack 1,n_{\rm seg} \rbrack}$ are the local Zernike coefficients on each segment, and $(\mathbf{r}k){k\in [1,n_{\rm seg}]}$ are the position vectors from the center of the pupil to the centers of the segments.

This expression can be separated into two factors: (1) a low-order envelope, $\left \Vert \widehat{Z}(\mathbf{u}) \right \Vert^2$, which depends on the segment shape and the Zernike polynomial composing the segment-level wavefront error \citep{Yaitskova2002, Yaitskova2003}, and (2) a sum of interference fringes between all pairs of segments, whose amplitudes depend on both the aberration magnitude and the coronagraph type. An alternative way to visualize this envelope is shown in Fig.~\ref{fig:Figure0_PASTIS}: the PSF of a single segment corresponds to the low-order envelope when the segment is replicated across the pupil (here with $7$ and $19$ segments), while the replication of the segment generates only the higher-frequency interference fringes. Equation~\ref{eq:PASTIS} generalizes this phenomenon to coronagraphic PSFs, in which the interference fringes are modulated by the contribution of each segment to the contrast.

In the presence of segment-level aberrations, the coronagraphic PSF is degraded proportionally to the low-order envelope, $\left \Vert \widehat{Z}(\mathbf{u}) \right \Vert^2$. For piston-only segment-level aberrations, this envelope corresponds to the PSF of a single segment and is strongest at small angular separations, with a characteristic radius of $1.22 \lambda/d$, where $d$ is the segment diameter. If the pupil contains $N$ segments across its diameter, so that $D = N \times d$, the low-order envelope then has a typical size of $1.22 \times N \lambda / D$.

\begin{figure}
   \begin{center}
   \begin{tabular}{c}
   \includegraphics[width=5cm]{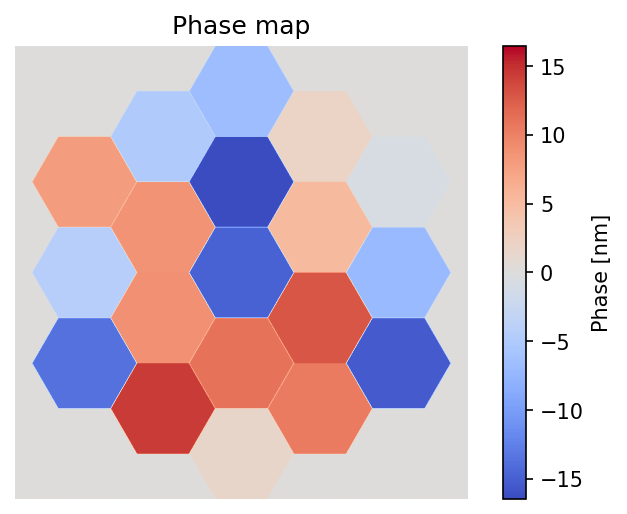} \\
   \includegraphics[width=8.5cm]{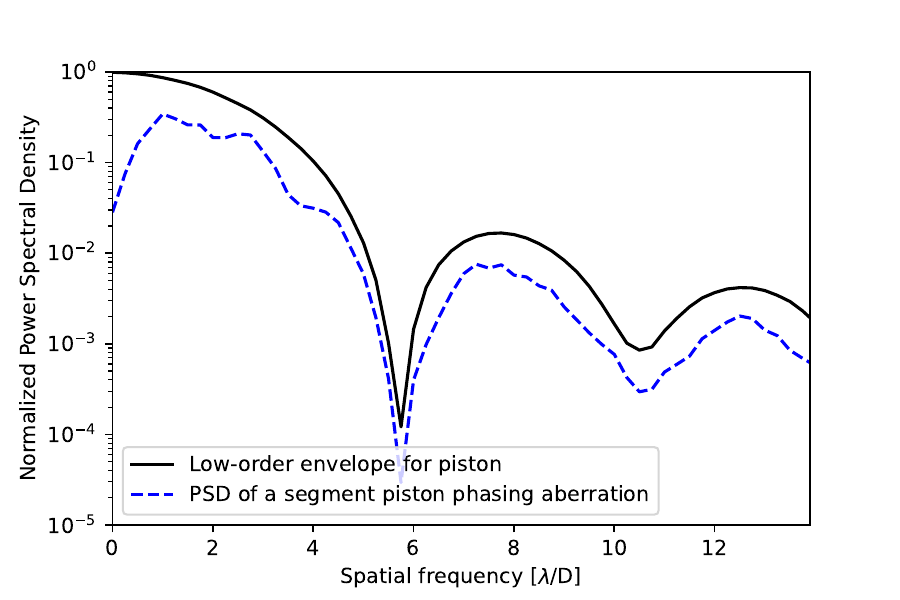}
   \end{tabular}
   \end{center}
   \caption[Fig] 
   { \label{fig:Figure1_DSP} Power spectral density (PSD) of segment-level piston aberrations: (top) example of a segment phasing error map; (bottom) PSD of this phase map along with the PSF of a single segment. The low-order envelope of the PSD corresponds to the segment PSF.}
   \end{figure}

For instance, Fig.~\ref{fig:Figure1_DSP} shows the low-order envelope of a single segment (black) together with the Power Spectral Distribution (PSD) of a random phasing error, for a segmented pupil with approximately $5$ segments across the diameter (precisely $4.7$). The first zero of the PSD occurs at $1.22 \times N \lambda / D$, so phasing errors at lower spatial frequencies have a correspondingly larger impact up to roughly $5 \lambda / D$.

The coronagraph, and in particular the FPM, acts as a high-pass filter with a cutoff frequency set by the FPM radius. This implies that if the inner working angle (IWA) is larger than roughly $N \lambda / D$, then most of the low-order envelope is blocked by the FPM, and phasing errors have a reduced impact on the coronagraphic PSF contrast floor (the impact on the Strehl ratio remains). Conversely, if the IWA is smaller than approximately $N \lambda / D$, then phasing errors significantly degrade the contrast at small angular separations.

\begin{figure}
   \begin{center}
   \begin{tabular}{c}
   \includegraphics[width=8.5cm]{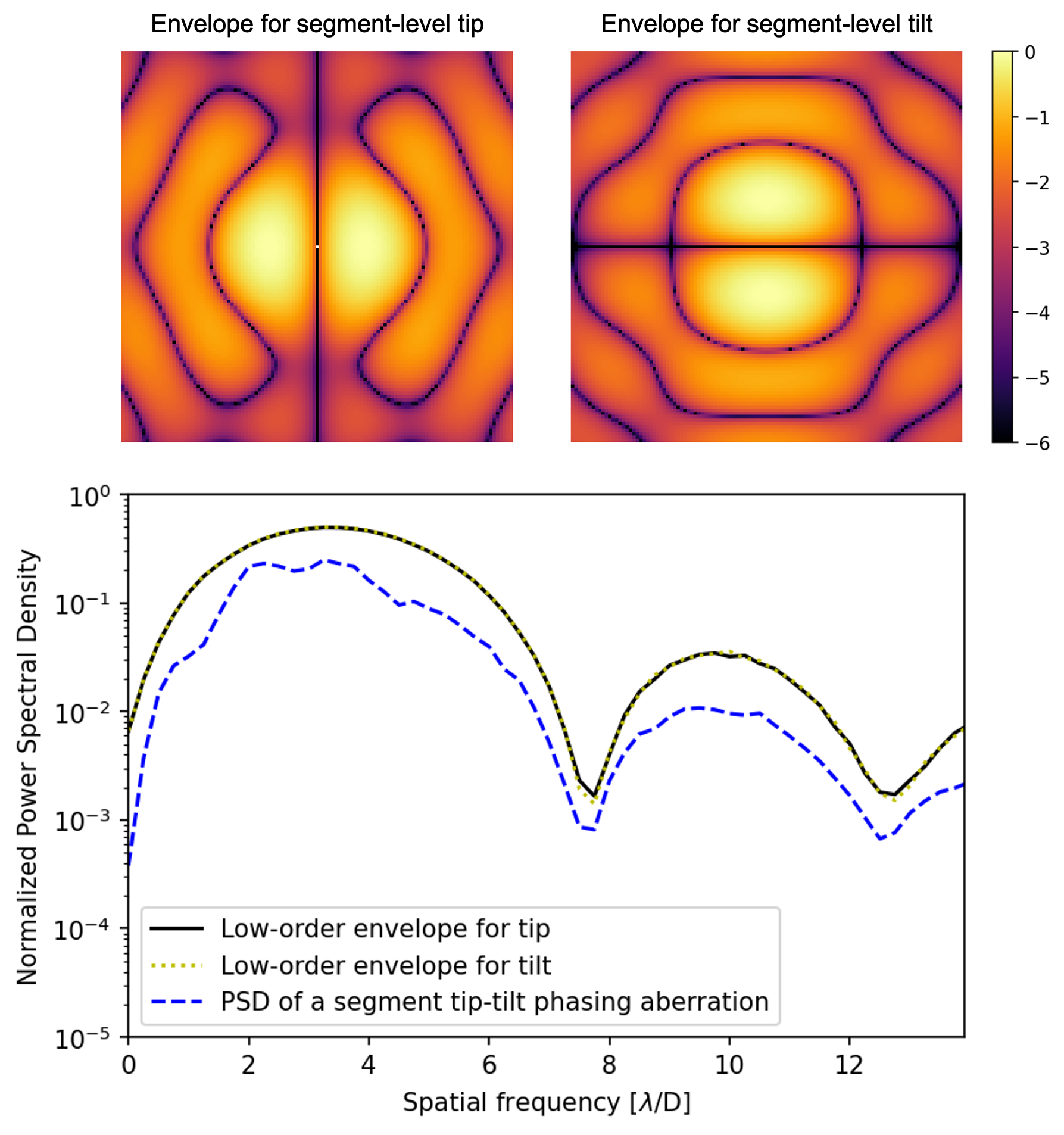}
   \end{tabular}
   \end{center}
   \caption[Fig] 
   { \label{fig:Figure2_TipTilt} PSD of segment-level tip-tilt aberrations: (top) 2D low-order envelopes for tip and tilt, (bottom) 1D azimuthally averaged low-order envelopes for tip and tilt, together with the PSD of a random segment-level tip-tilt aberration, contained within these envelopes.}
   \end{figure}

Segment phasing errors also generate local tip and tilt, which produce different low-order envelopes than segment-level pistons. Fig.~\ref{fig:Figure2_TipTilt} shows these low-order envelopes and their azimuthal averages, along with the PSD of a random tip-tilt phasing error of $1$ nm rms. While the low-order piston envelope has the highest impact below $5 \lambda/D$, the low-order tip and tilt envelopes are higher between $1$ and $6 \lambda/D$. Consequently, the dark region, if not covered by the FPM, will be more affected by segment-level tip and tilt errors at these separations.

\subsection{General implications of phasing errors for low-order wavefront sensing}

Using the light rejected by the coronagraph, for example, by the focal plane mask or by the Lyot stop, to measure for the incoming wavefront aberrations presents two main advantages. Not only does it measure the aberrations in planes close to the coronagraphic planes that are very sensitive to aberrations, reducing non-common path aberrations with the wavefront sensor, but it also makes use of photons not used by the coronagraphic path. 

However, working in the light rejected by the coronagraph involves using a filtered signal. In the case of an APLC, with a low-order wavefront sensor using the light rejected by the center of the focal plane mask, only the core of the PSF is retained in the wavefront sensing arm. The FPM therefore acts as a spatial low-pass filter, dampening the spatial frequencies above the cutoff frequency defined by its diameter. Even if the segment-level aberrations spatial frequency components extend to frequencies above the FPM cutoff frequencies, as presented in Fig.~\ref{fig:Figure1_DSP} and Fig.~\ref{fig:Figure2_TipTilt}, they also have content at lower frequencies. This means there is some signal for the measurement of the segment-level aberrations, that depends on the FPM diameter, and on the density of segments. Depending on the structure of the input pupil, apodizer, and FPM, reconstruction of the input aberration might introduce confusion between the input modes, which requires more photons to properly disentangle the signals.  

To approximate the ability to reconstruct the phasing errors of the primary aperture, we calculate the sensitivity of the wavefront sensor to photon noise, $s_\gamma$, as defined by \cite{Chambouleyron2021}. Using the same notations, the sensitivity to photon noise $s_\gamma$ to a mode $\phi_i$ can be computed using the relation: $$s_\gamma(\phi_i) = \left|\left|\frac{\delta I(\phi_i)}{\sqrt{I_0}} \right|\right|_2 ,$$ where $||\cdot||$ is the L2 norm, $I_0$ is the intensity on the wavefront sensor in the absence of aberrations and $\delta I(\phi_i)$ is the difference of the positive poke and the negative poke measured on the wavefront sensor, normalized by twice the phase standard deviation, in radians, of a single poke. $\delta I(\phi_i)$ is usually referred to as the ``double difference`` and is frequently used as a column of the interaction matrix, recording the impact of $\phi_i$.

\subsection{Zernike Wavefront sensor as a low-order wavefront sensor}

The Zernike wavefront sensor (ZWFS) \citep{N'Diaye2013, Bloemhof2003, Dohlen2004} is a highly sensitive wavefront sensing technique \citep{Chambouleyron2021, Guyon2005}, well suited to the stringent constraints of space instrumentation \citep{Ruane2020, Steeves2020}. It has already been deployed on instruments such as VLT/SPHERE \citep{Vigan2019, N'Diaye2016}. One can also note the use of a ZWFS to detect segment phasing errors at the Keck Observatory \citep{Salama2024}.

In the context of coronagraphic systems, the FPM can be used not only to block the on-axis starlight while transmitting the off-axis source photons, but also to reflect this starlight, which carries low-order aberration information toward a dedicated ZWFS \citep{Pourcelot2022, Pourcelot2023}. This approach has been experimentally validated in the context of the Nancy Grace Roman Space Telescope, on which it will be implemented \citep{Shi2016}. However, in the case of segment-level phasing errors, the FPM low-order filtering has a direct impact on the wavefront reconstruction performed by the ZWFS.

With a Zernike dimple or diameter of $1.06\lambda/D$, sometimes called ``Classical ZWFS``, and without spatial filtering by a FPM, the sensitivity $s_\gamma$ is constant with the spatial frequency, around 1.2, excepted for the very low spatial frequencies around 1 cycle / pupil where it drops, because the aberration information lies within the dimple of the ZWFS mask and therefore does not get encoded in the interferences. In our situation, two elements cause the sensitivity value to be lower. The first one is the filtering by the focal plane mask that strongly attenuates the high spatial frequencies, reducing the ability of any sensor to measure them; the second one is the apodizer that also changes the spatial frequency content of the input pupil and that throws away a percentage of photons, especially for the segments that are the most masked, that tend to be the outer ones. For a fair comparison of the sensitivity between apodizations, we normalize the measurement by the power in the entrance pupil, without the apodizer, as a loss of photons translates into a loss of robustness to photon noise of the wavefront sensor.  

Finally, we used the ``Classical ZWFS`` for the simulations with a dot diameter of $1.06\lambda/D$, which yields the highest sensitivity for the segment aberrations that are above 1 cycle/pupil, as developed in \citep{Chambouleyron2021}. However, rather than providing absolute values of sensitivities, we expect the ratio between the configurations to not be dependent of the ZWFS diameter. As a result, the configurations yielding better sensitivities are expected to do so with for example a ZWFS with a dot diameter of $2\lambda/D$, keeping this comparative study valid. Going for larger values would, however, degrade the performance at lower spatial frequencies and have an impact on, for example, configurations with fewer segments, but the detailed values are beyond the scope of this work. 

\section{Impact of the focal-plane mask radius and system cutoff frequency}
\label{s:Impact of the focal-plane mask radius and system cutoff frequency}

This section examines the effect of the FPM radius on the robustness of coronagraphic performance to segment phasing errors, for a given segment density. Two studies are considered: the impact of segment phasing aberrations on the coronagraphic image and performance, and the ability of the low-order ZWFS to reconstruct these aberrations.

\subsection{Numerical experiment: setup and parameter samples}

\begin{figure*}
   \begin{center}
   \begin{tabular}{c}
   \includegraphics[width=17.5cm]{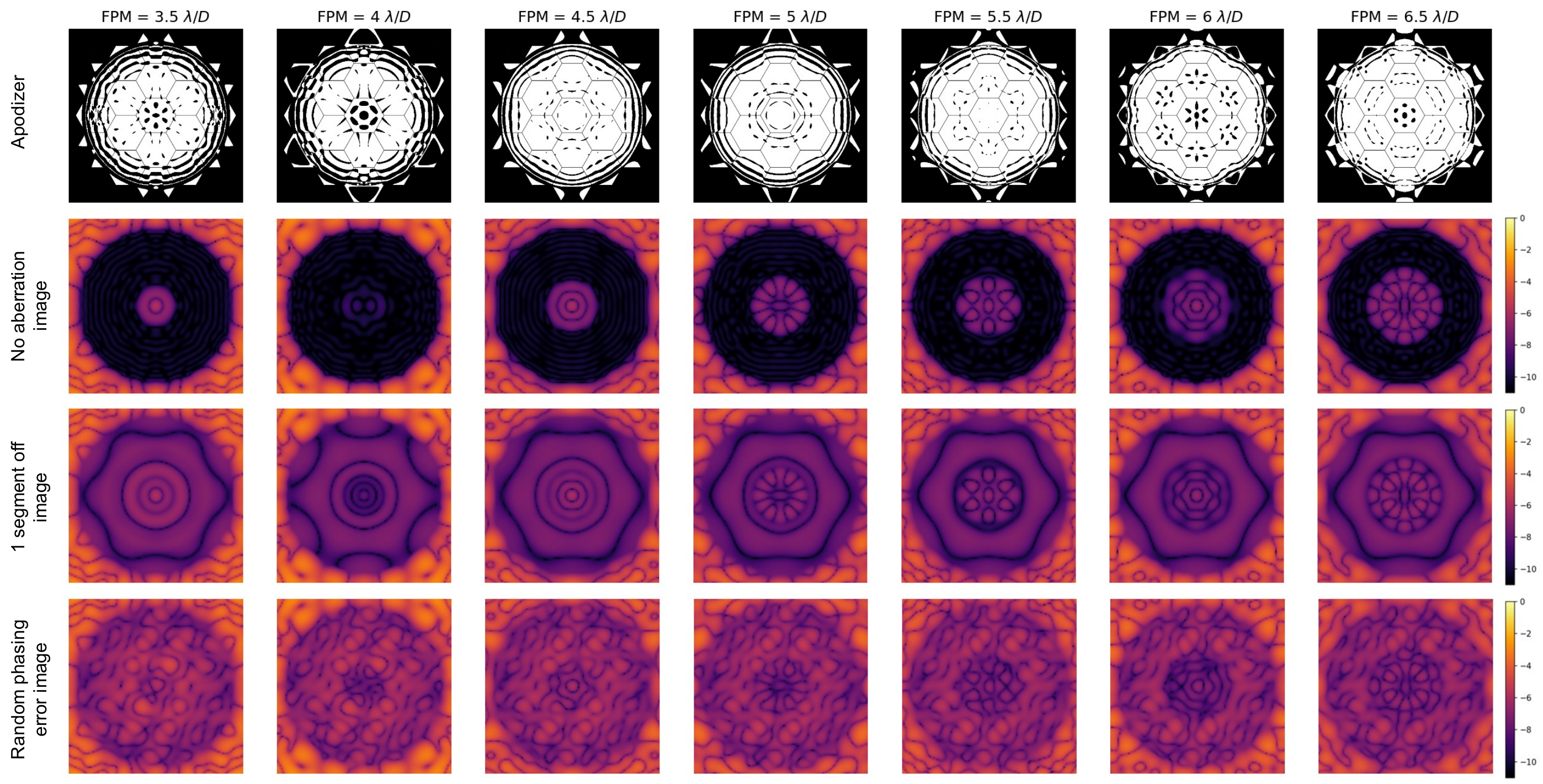}
   \end{tabular}
   \end{center}
   \caption[Fig] 
   { \label{fig:Figure4_Coronos} The seven APLC designs and images, with FPM radii ranging from (left) $3.5$ to (right) $6.5 \lambda/D$: (row 1) apodizer mask solutions, (row 2) coronagraphic PSFs without aberrations, (row 3) coronagraphic PSFs with the central segment displaced by 1 nm, (row 4) coronagraphic PSFs with a 1 nm rms random phasing error.}
   \end{figure*}

\begin{figure}
   \begin{center}
   \begin{tabular}{c}
   \includegraphics[width=4.5cm]{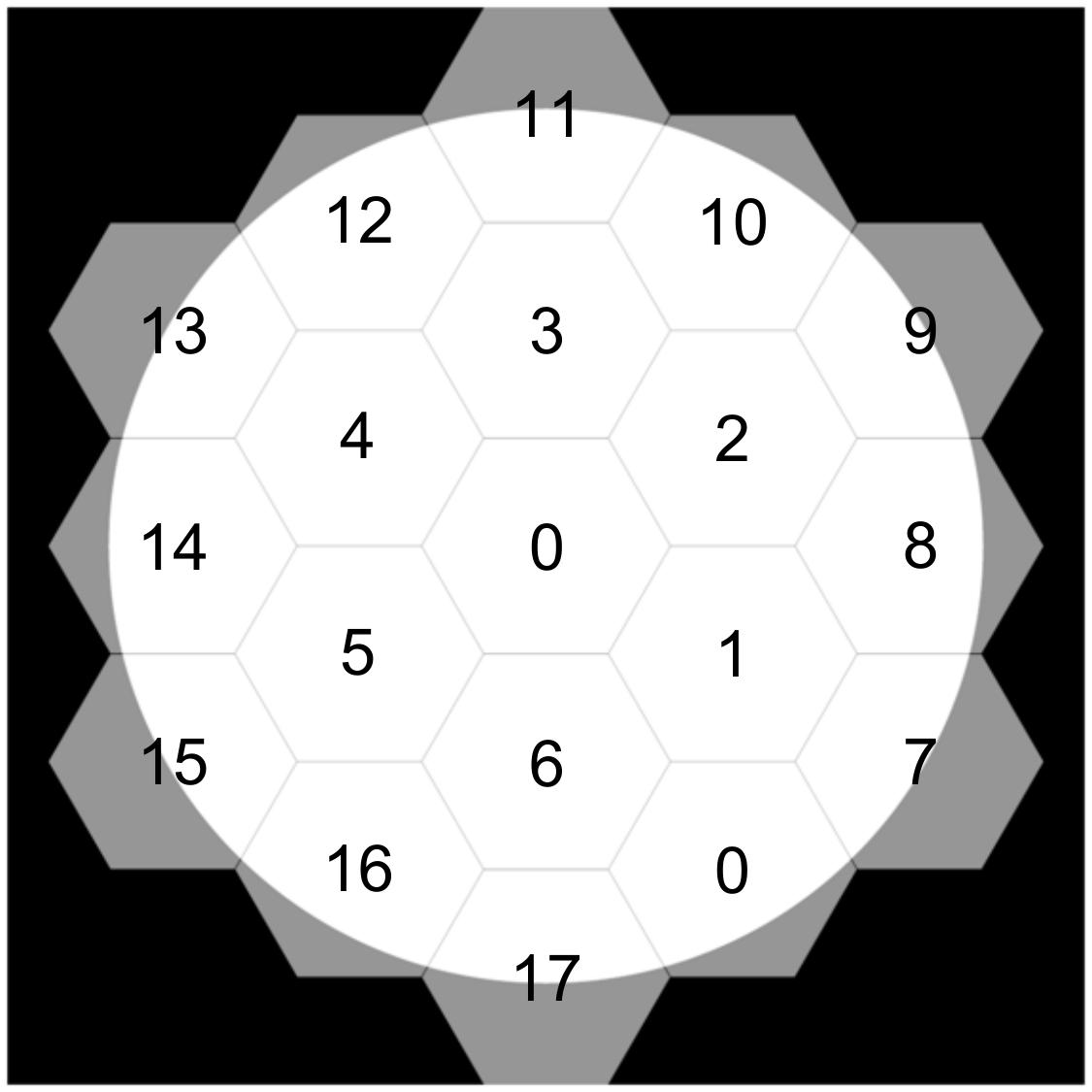}
   \end{tabular}
   \end{center}
   \caption[Fig] 
   { \label{fig:Figure3_LyotStop} Schematic of the Lyot stop, here visually superimposed on the 19-segment entrance pupil. Segment numbering is indicated, as referenced for the modes ordering of Section \ref{s:Phasing error sensing and mitigation}.}
   \end{figure}

For this study, we consider a segmented pupil with a $6$m inscribed diameter ($7.26$m external diameter) and $5$ hexagonal segments across its diameter, for a total of $19$ segments. We design seven different APLCs, each made of three amplitude masks: 

$\bullet$ an apodizer, generated following the optimization algorithm of \cite{Nickson2022} and shown in Fig. \ref{fig:Figure4_Coronos} (top row),

$\bullet$ a FPM, with a radius ranging from $3.5$ to $6.5 \lambda/D$,

$\bullet$ a circular Lyot stop, slightly smaller than the entrance pupil (see Fig. \ref{fig:Figure3_LyotStop}) and identical for all seven APLCs.

For these design optimizations, we use APLC-Optimization, a software toolkit developed by the SCDA (Segmented Coronagraph Design \& Analysis) research team at the Space Telescope Science Institute (STScI) \citep{Nickson2022}. In addition, all numerical propagations in this study are performed with the HCIPy software package \citep{Por2026}. 

For this first study, the specifications of the seven outcomes coronagraphs are listed in Table~\ref{table:Specifications}. In the absence of aberrations, they achieve contrast floors below $10^{-10}$ within their respective dark zones, as shown in Fig.~\ref{fig:Figure4_Coronos} (second row).

\begin{table*}[h!]
\centering
\begin{tabular}{|c|c|c|c|c|c|}
  \hline
  \textbf{Case} & \textbf{Contrast} & \textbf{IWA ($\lambda/D$)} & \textbf{OWA ($\lambda/D$)} & \textbf{FPM radius} ($\lambda/D$) & \textbf{Transmission} \\
  \hline
  \hline
  1 & $4.3 \times 10^{-11}$ & $3.4$ & $12$ & $3.5$ & $62\%$ \\
  2 & $3.2 \times 10^{-11}$ & $3.9$ & $12$ & $4.0$ & $62\%$ \\
  3 & $4.6 \times 10^{-11}$ & $4.4$ & $12$ & $4.5$ & $73\%$ \\
  4 & $4.1 \times 10^{-11}$ & $4.9$ & $12$ & $5.0$ & $70\%$ \\
  5 & $3.1 \times 10^{-11}$ & $5.4$ & $12$ & $5.5$ & $75\%$ \\
  6 & $3.7 \times 10^{-11}$ & $5.9$ & $12$ & $6.0$ & $77\%$ \\
  7 & $3.9 \times 10^{-11}$ & $6.4$ & $12$ & $6.5$ & $82\%$ \\
  \hline
\end{tabular}
\caption{Specifications of the seven APLC designs. In all cases, the primary mirror consists of 19 hexagonal segments, and only the FPM radius and the IWA are varied in the design optimization requirements.}
\label{table:Specifications}
\end{table*} 

\subsection{Passive robustness by optical design}

\subsubsection{Robustness to segment-level piston}

Of the seven APLCs introduced in the previous section, we study their robustness to a wide range of segment-level piston phasing errors.

In Fig.~\ref{fig:Figure4_Coronos}, the third and fourth rows show the coronagraphic PSFs, respectively, when a single segment is displaced by $1$ nm and when a random piston-only phasing error of $1$ nm rms is applied to the entrance mirror. In the third row, the first zero of the segment low-order envelope exceeds the IWA for the first three cases (FPM radii from $3.5$ to $4.5 \lambda/D$), indicating that phasing errors have a strong impact in this region of the dark hole. In the last four cases (FPM radii from $5$ to $6.5 \lambda/D$), most of the low-order envelope is blocked by the FPM, so the impact of phasing errors on the contrast is reduced.

\begin{figure}
   \begin{center}
   \begin{tabular}{c}
   \includegraphics[width=8.5cm]{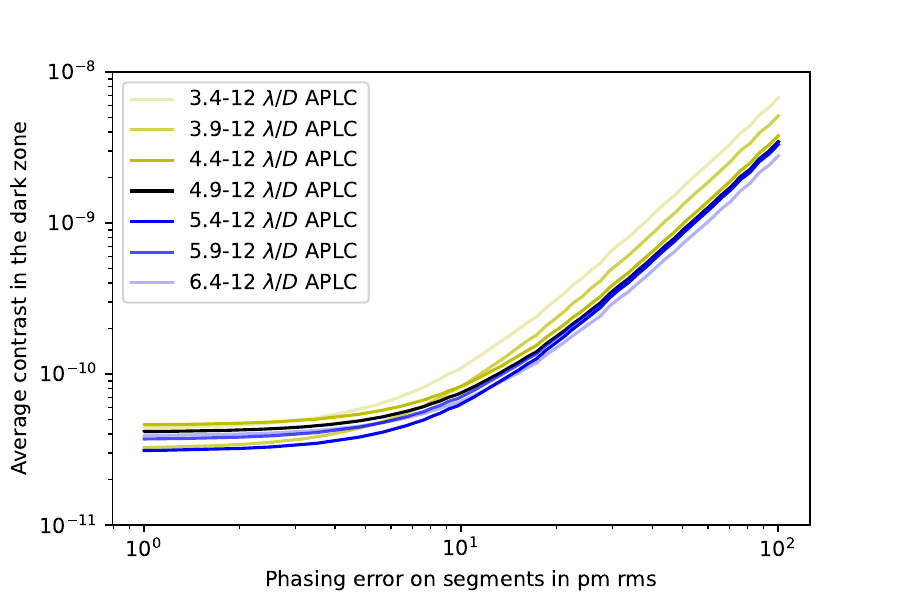} \\
   \includegraphics[width=8.5cm]{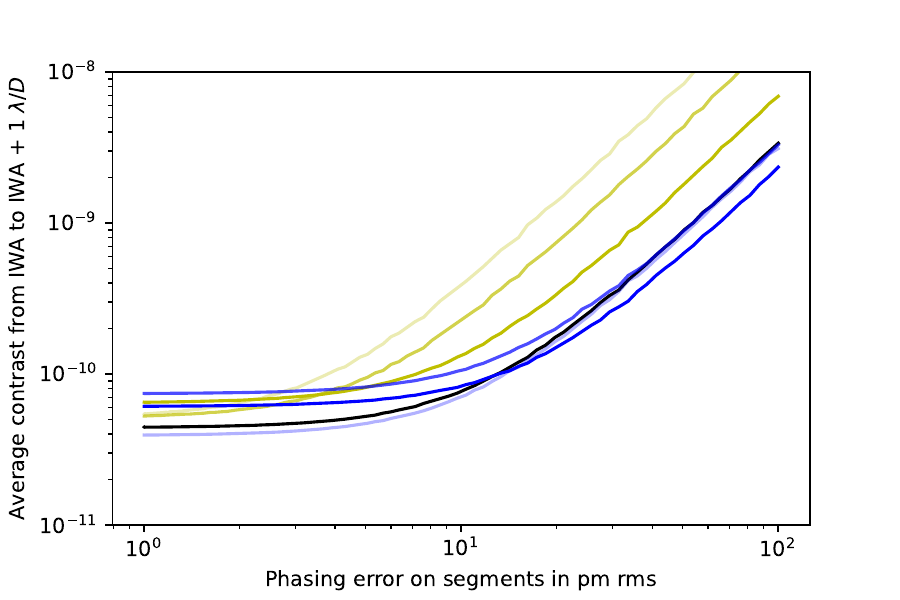}
   \end{tabular}
   \end{center}
   \caption[Fig] 
   { \label{fig:Figure5_HockeyCross} Comparison of the robustness of the seven designs to segment-level piston phasing errors: (top) average contrast in the dark region as a function of the wavefront error amplitude; (bottom) average contrast near the IWA as a function of the wavefront error amplitude.}
   \end{figure}

To quantify this impact relative to the FPM diameter, we perform end-to-end simulations of the seven coronagraphic systems, generating high-contrast images with increasing segment-level piston errors. The simulations were performed using a classical end-to-end propagation algorithm, with $200$ amplitude steps from $1$ to $100$ pm rms, and for each amplitude step, $200$ random segment phasing errors were generated and propagated through the system to compute $200$ coronagraphic PSFs. Figure~\ref{fig:Figure5_HockeyCross} shows the resulting performance, i.e., the mean values of the $200$ contrasts of these PSFs as a function of the amplitude step: (top) the effect of increasing phasing errors on the average contrast in the dark region, and (bottom) their effect on the contrast at small angular separations, from the IWA to IWA$+1 \lambda/D$. In the first plot, APLCs with FPM radii larger than $5 \lambda/D$ (i.e., system cutoff frequency) appear slightly less sensitive than the others; for a target contrast of $10^{-10}$, for instance, the phasing constraints are relaxed by a factor of $1.6$. In the second plot, the impact of the low-order envelope is even more pronounced: for the same target contrast, the phasing constraints are relaxed by a factor of $3.5$. 

\subsubsection{Robustness to segment-level tip–tilt}

\begin{figure}
   \begin{center}
   \begin{tabular}{c}
   \includegraphics[width=8.5cm]{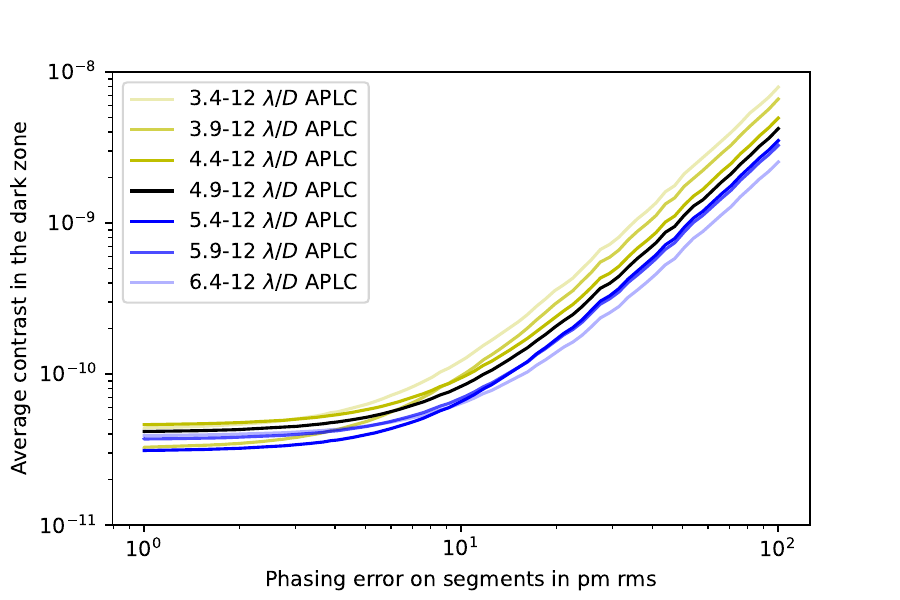}\\
   \includegraphics[width=8.5cm]{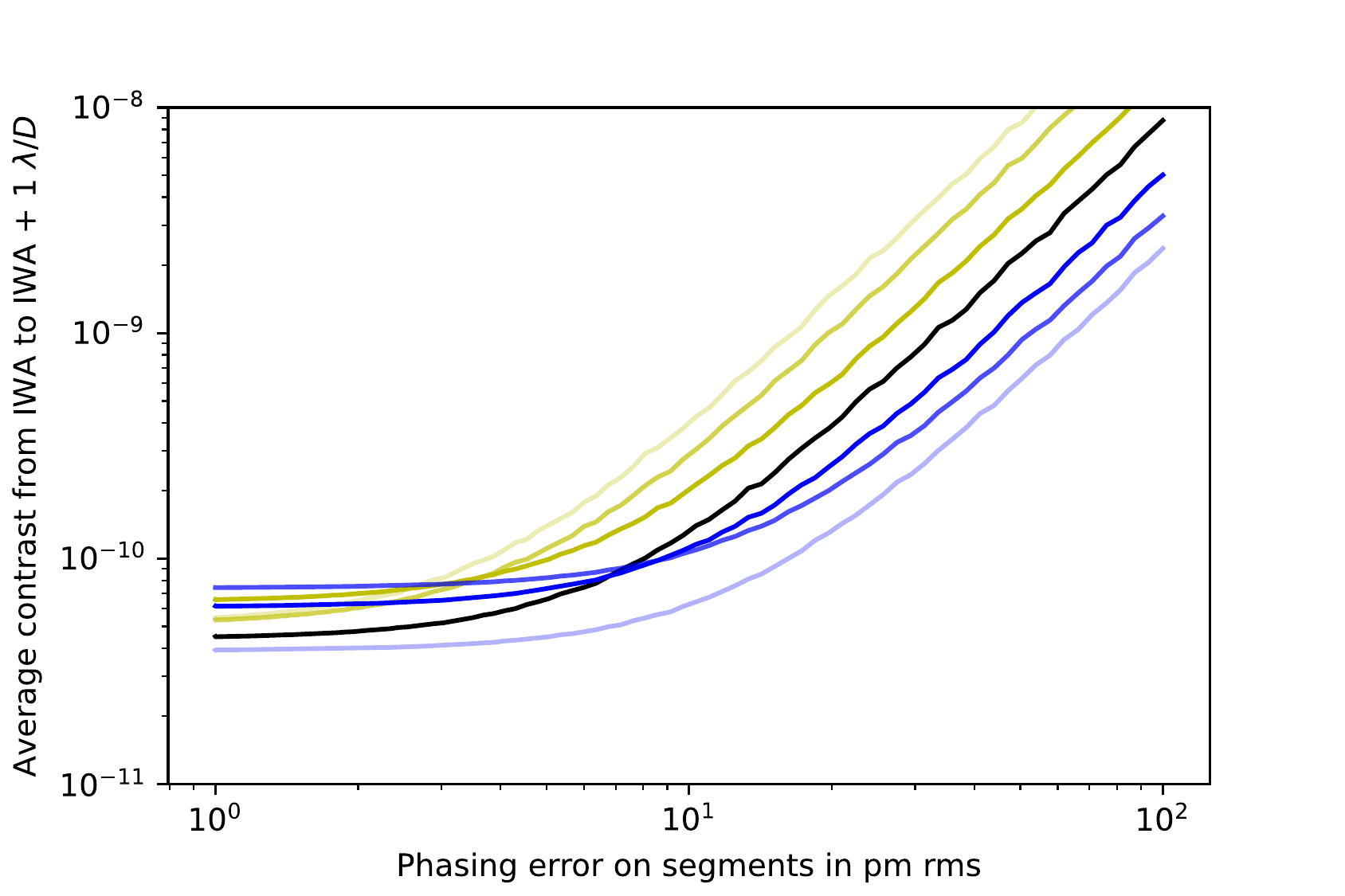}
   \end{tabular}
   \end{center}
   \caption[Fig] 
   { \label{fig:Figure7_HockeyCross_TipTilt} Comparison of the robustness of the seven designs to segment-level tip-tilt phasing errors: (top) average contrast in the dark region as a function of wavefront error amplitude; (bottom) average contrast near the IWA as a function of wavefront error amplitude.}
   \end{figure}

As mentioned above (see Fig.~\ref{fig:Figure2_TipTilt}), the low-order envelopes for segment-level tips and tilts are higher between $1$ and $6 \lambda/D$. This is verified by propagating increasing segment-level tip-tilt errors through end-to-end simulations of the seven coronagraphic systems. The simulations were performed using a classical end-to-end propagation algorithm, with $200$ amplitude steps ranging from $1$ to $100$ pm rms, and for each amplitude step, $200$ random segment phasing errors were generated and propagated through the system to compute $200$ coronagraphic PSFs. The resulting error budget is shown in Fig.~\ref{fig:Figure7_HockeyCross_TipTilt} (the mean values of the $200$ contrasts of these PSFs as a function of the amplitude step) with: (top) the impact of increasing errors on the average contrast in the dark region, and (bottom) their impact on the contrast at small angular separations, from IWA to IWA$+1 \lambda/D$. Overall, the higher the FPM radius, the more robust the APLC is to segment-level tip-tilt errors. Quantitatively, in our cases, increasing the FPM radius from $3.5 \lambda/D$ to $6.5 \lambda/D$ relaxes the tip-tilt constraints by a factor of $1.8$ (top plot), and by a factor of $4.0$ in the $1 \lambda/D$ large separations (bottom plot).

As a conclusion, to passively stabilize the contrast at small angular separations, a first relaxation of the phasing constraints can be achieved by having a number of segments across the pupil diameter, $N$, smaller than the target IWA in units of $\lambda/D$. For an IWA as small as $3 \lambda/D$, this corresponds to a segmented mirror with no more than three segments across the pupil diameter (seven segments in total). The impact of segment density on robustness to segment-level phasing errors is discussed in Section~\ref{s:Impact of segment density}.

\subsection{Phasing error sensing with a low-order wavefront sensor}
\label{s:Phasing error sensing and mitigation}

\begin{figure*}
    \centering
    \includegraphics[width=17.5cm]{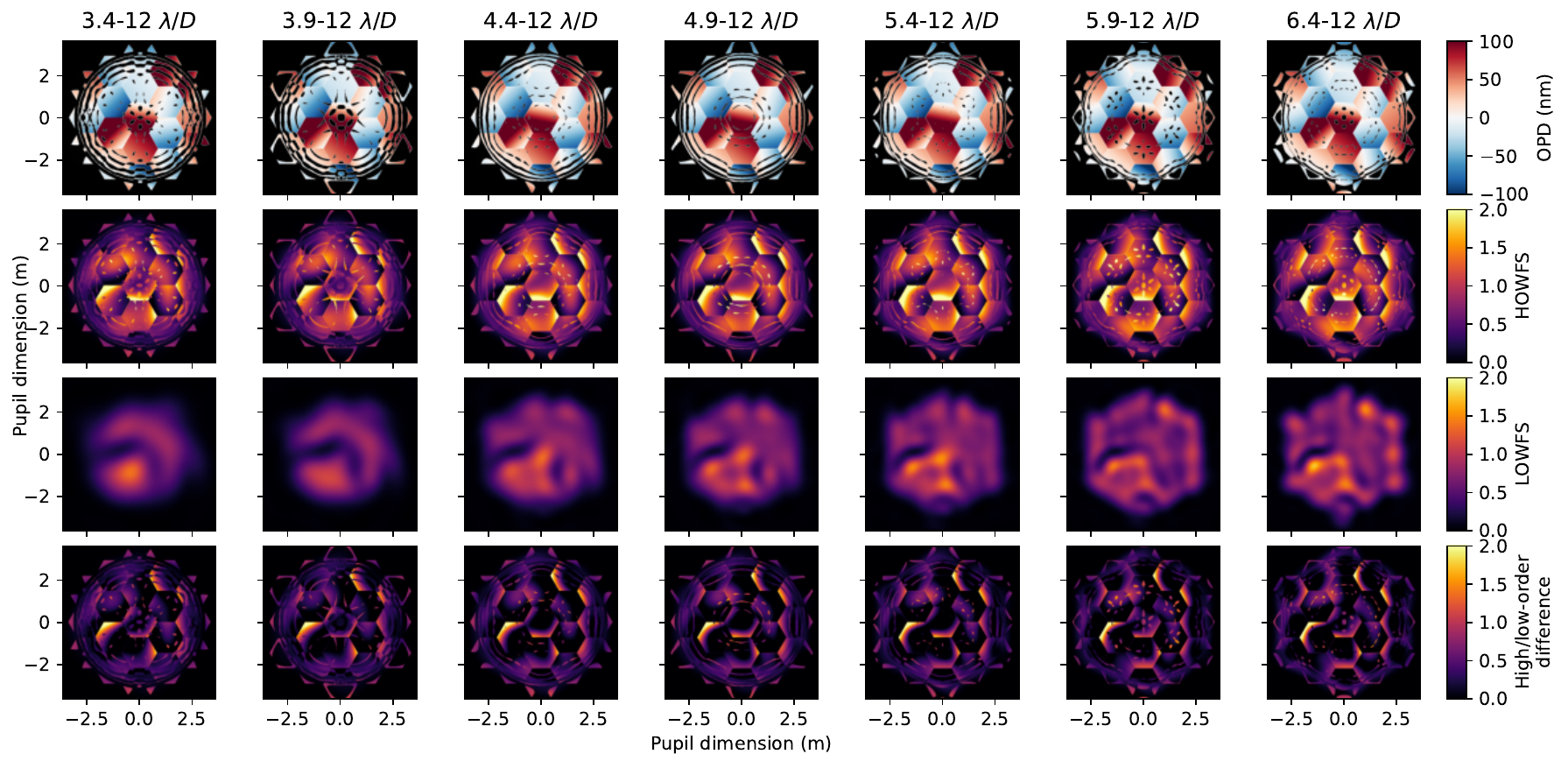}
    \caption{For the $7$ APLC cases, (row 1) input wavefront error (Here $50$nm rms of segment-level piston-tip-tilt. These values are chosen artificially high for the ease of the visualization of the difference between the difference cases) and intensities of (row 2) their reconstructions through a high-order ZWFS (no filtering), (row 3) low-order ZWFS reconstructions, i.e., filtered by the FPM before reaching the ZWFS, (row 4) the difference of row 2 minus row 3.}
    \label{fig:Figure8_intensities}
\end{figure*}

\begin{figure}
    \centering
    \includegraphics[width=8.5cm]{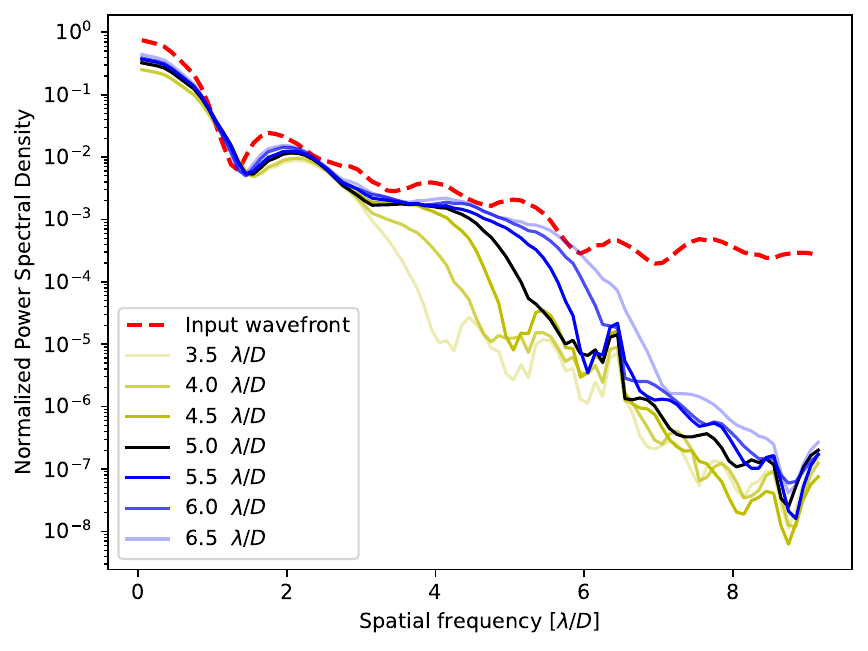}
    \caption{Power spectral densities of the reconstructed aberrations for different FPM radii ($3.5$–$6.5 \lambda/D$), compared to that of the input aberrated electric field.}
    \label{fig:Figure8_psds}
\end{figure}

For a specific phasing error, Figure~\ref{fig:Figure8_intensities} illustrates the impact of the focal plane mask (FPM) low-order spatial filtering on the ZWFS wavefront reconstruction for the seven configurations of APLCs (here, mostly the FPM diameter matters, ranging from $3.5$ to $6.5 \lambda/D$). The first row shows the input error phase map, with amplitudes of $50$ nm rms. The second and third rows respectively display the intensity maps reconstructed by an ideal high-order sensor (HOWFS, no filtering by any FPM) and a low-order ZWFS (LOWFS) with varying FPM radii, and the fourth row shows the difference between the two reconstructions. The LOWFS reconstruction is strongly limited by the FPM filtering, retaining only smooth, large-scale wavefront variations and missing the high-spatial-frequency phasing errors. This is further quantified by the PSD analysis shown in Figure~\ref{fig:Figure8_psds}, where the reconstructed wavefront PSDs progressively deviate from the input PSD starting at approximately the FPM cutoff frequency. 

\begin{figure}
    \centering
    \includegraphics[width=8.5cm]{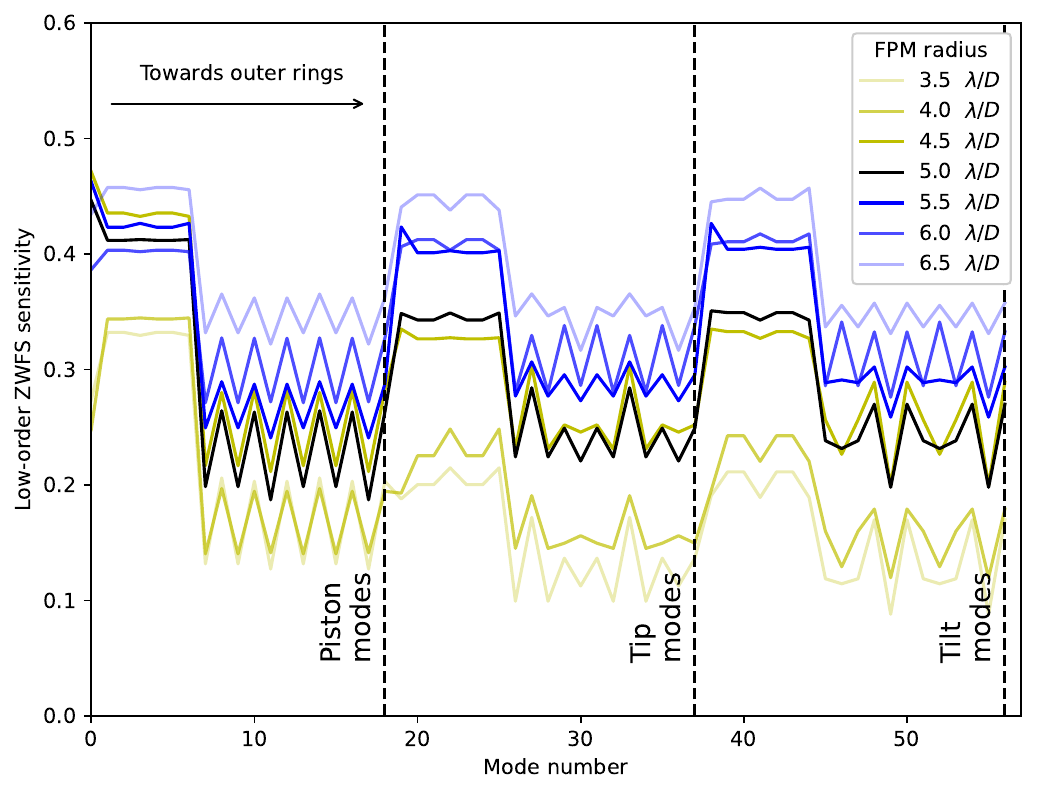}
    \caption{Sensitivity of the low-order ZWFS to segment piston, tip, and tilt modes as a function of mode number, for the seven FPM radii. Within each mode family, the mode index increases from the innermost to the outermost segment ring. }
    \label{fig:Figure8_sensitivities}
\end{figure}

On a more general consideration, Figure~\ref{fig:Figure8_sensitivities} shows the low-order ZWFS sensitivity as a function of mode number, for the seven FPM radii considered in this study. The modes are organized into three Zernike polynomials (segment piston, tip, and tilt) separated by vertical dashed lines, with mode indices increasing from inner to outer segment rings within each group (see also the numbering of Fig.\ref{fig:Figure3_LyotStop}). Two trends emerge consistently across all three segment-level Zernike polynomials. The sensitivity drops sharply from the innermost to the outer-ring modes, that is due to the apodizer pattern, that tend to hide the outermost segments to make the aperture more circular. Additionally, a systematic dependence on the FPM radius is observed: larger FPM radii (blue tones) yield higher ZWFS sensitivity across all modes, as a wider mask reflects more starlight toward the low-order sensing channel, while smaller radii (yellow tones) result in reduced sensitivity. On average over all modes, the sensitivity coefficients increase from $0.178$ to $0.384$ when the FPM radius goes from $3.5$ to $6.5\lambda/D$. This means that the wavefront is twice as sensitive with the largest FPM on the same segmentation modes.



\section{Impact of segment density}
\label{s:Impact of segment density}

From the previous section:

$\bullet$ the coronagraph is passively more robust to segment-level piston, tip, and tilt misalignments if the FPM radius is equal to or larger than $N \lambda/D$, where $N$ is the number of segments across the pupil diameter,

$\bullet$ the light reflected by the FPM can be efficiently used to reconstruct segment-level phasing errors with a Zernike wavefront sensor if the FPM is sufficiently large, typically equal to or larger than $N \lambda/D$,

For a target IWA around $3 \lambda/D$, this implies that the optimal primary mirror should be either monolithic or segmented with no more than three segments across its diameter (seven segments in total in the pupil). The aim of this section is to study the robustness of the instrument as a function of segment density, for given IWA, OWA, and FPM radius.

\subsection{Numerical experiment: setup and parameter samples}

\begin{figure*}
   \begin{center}
   \begin{tabular}{c}
   \includegraphics[width=14cm]{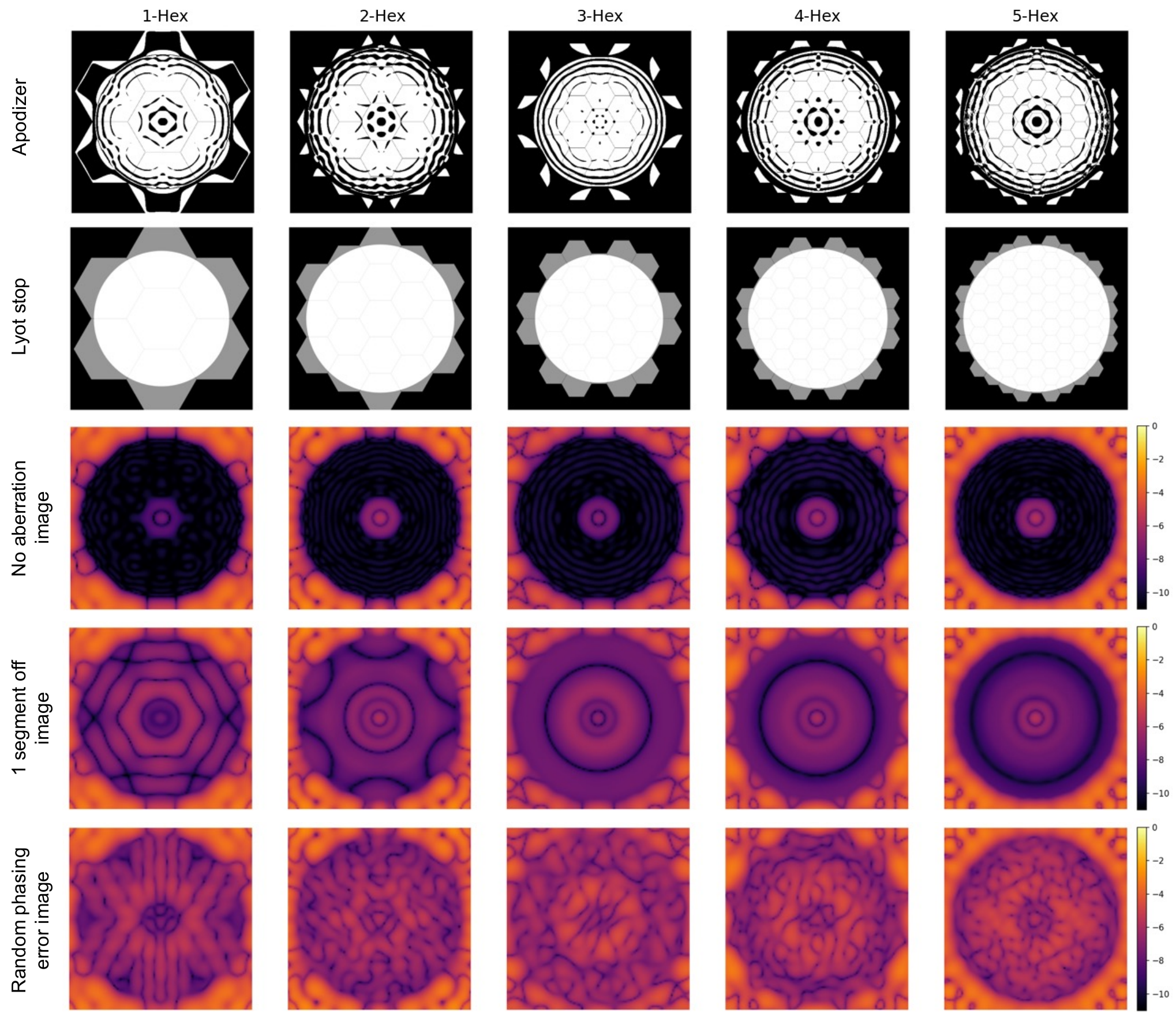}
   \end{tabular}
   \end{center}
   \caption[Fig] 
   { \label{fig:SEC4_Figure1_PSF} The five designs and their images, with segmentation types ranging from (left) $3$ to (right) $11$ segments across the pupil diameter: (row 1) apodizer mask solutions, (row 2) primary mirrors and Lyot stops, (row 3) coronagraphic PSFs without aberrations, (row 4) coronagraphic PSFs with the central segment displaced by $1$ nm, (row 5) coronagraphic PSFs with a $1$ nm rms random phasing error.}
   \end{figure*}

Five primary mirror configurations are compared, with 3 to 11 segments across the pupil, corresponding to 7, 19, 31, 55, and 85 segments in total (see Fig.~\ref{fig:SEC4_Figure1_PSF}, second row). They are labeled according to the number of hexagonal segment rings, from 1-Hex to 5-Hex, with 2-Hex having been introduced in the previous section.

From these five architectures, five APLCs are designed to achieve a contrast below $10^{-10}$ between $3.4$ and $12 \lambda/D$. The segment density of the entrance pupil varies across the designs, affecting both the apodizer and the Lyot stop (see Fig.~\ref{fig:SEC4_Figure1_PSF}, first and second rows). The resulting specifications are listed in Table~\ref{table:Specifications2}.

\begin{table*}[h!]
\centering
\begin{tabular}{|c|c|c|c|c|c|c|}
  \hline
  \textbf{Case} & \textbf{Nb of segments} & \textbf{Contrast} & \textbf{IWA ($\lambda/D$)} & \textbf{OWA ($\lambda/D$)} & \textbf{FPM radius} ($\lambda/D$) & \textbf{Transmission} \\
  \hline
  \hline
  1-Hex & 7  & $3.0 \times 10^{-11}$ & $3.4$ & $12$ & $3.5$ & $66\%$ \\
  2-Hex & 19 & $4.1 \times 10^{-11}$ & $3.4$ & $12$ & $3.5$ & $64\%$ \\
  3-Hex & 31 & $5.0 \times 10^{-11}$ & $3.4$ & $12$ & $3.5$ & $70\%$ \\
  4-Hex & 55 & $9.1 \times 10^{-11}$ & $3.4$ & $12$ & $3.5$ & $71\%$ \\
  5-Hex & 85 & $3.7 \times 10^{-11}$ & $3.4$ & $12$ & $3.5$ & $69\%$ \\
  \hline
\end{tabular}
\caption{Specifications of the five APLC designs. Only the primary mirror segmentation type and the Lyot stop are varied in the design optimization requirements.}
\label{table:Specifications2}
\end{table*}

The five apodizers and the five Lyot stops (superimposed on the pupil) are shown in Fig.~\ref{fig:SEC4_Figure1_PSF}. This figure also displays the coronagraphic PSFs without aberrations, as well as the coronagraphic PSFs when a single segment is displaced by $1$ nm and under a $1$ nm rms random phasing error on the primary mirror.

\subsection{Passive robustness by optical design}

\subsubsection{Robustness to segment-level piston}

\begin{figure}
   \begin{center}
   \begin{tabular}{c}
   \includegraphics[width=8.5cm]{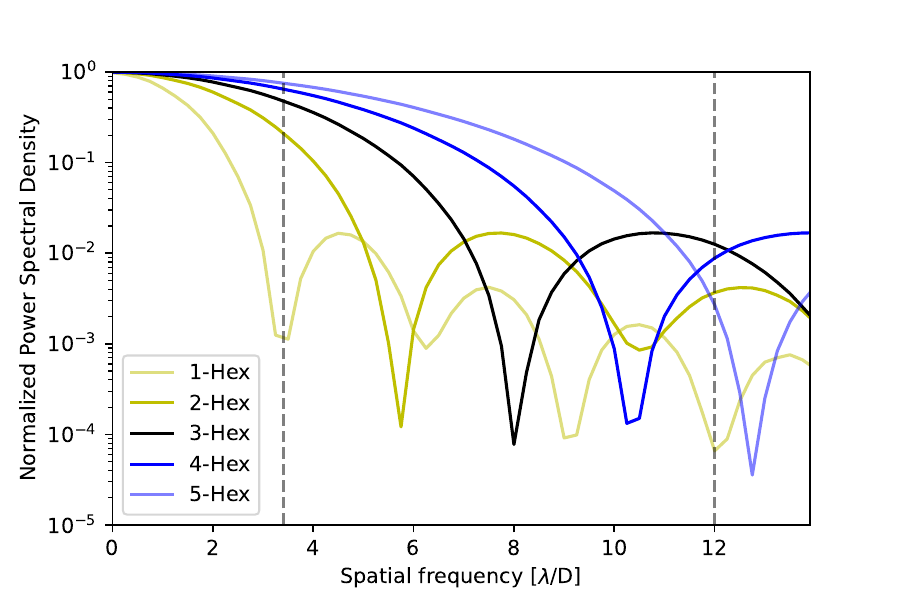}\\
   \includegraphics[width=8.5cm]{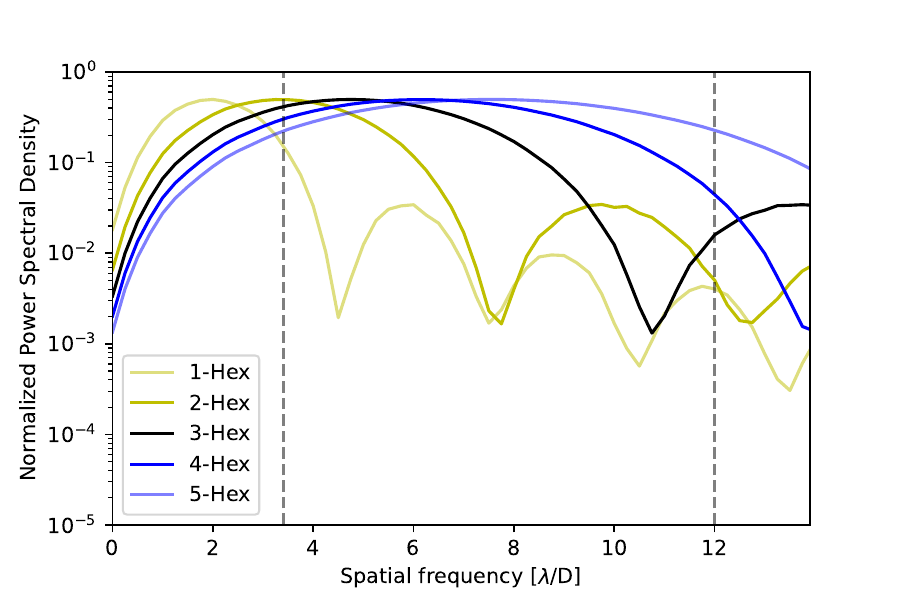}
   \end{tabular}
   \end{center}
   \caption[Fig] 
   { \label{fig:SEC4_Figure_DSP} Normalized low-order envelopes for (top) segment-level piston errors, (bottom) segment-level tip errors, each for the five segment densities. The gray dashed vertical lines indicate the dark-zone edges.}
   \end{figure}

The envelope of the PSD associated with a segment-level piston error depends on the size of the segment relative to the pupil diameter. Figure~\ref{fig:SEC4_Figure_DSP} (top) shows the five envelopes corresponding to the different segmentation types, together with the dark-zone edges. The envelopes reach their maximum between $0$ and $N \lambda/D$, where $N$ is the number of segments across the pupil diameter. Overall, the envelope of the 1-Hex configuration (three segments across the pupil) is the only one that appears lower than the other envelopes within the entire dark region, particularly at small angular separations, so close to the IWA ($3.4 \lambda/D$), since its low-order peak is cut off by the FPM. 

\begin{figure}
   \begin{center}
   \begin{tabular}{c}
   \includegraphics[width=8.5cm]{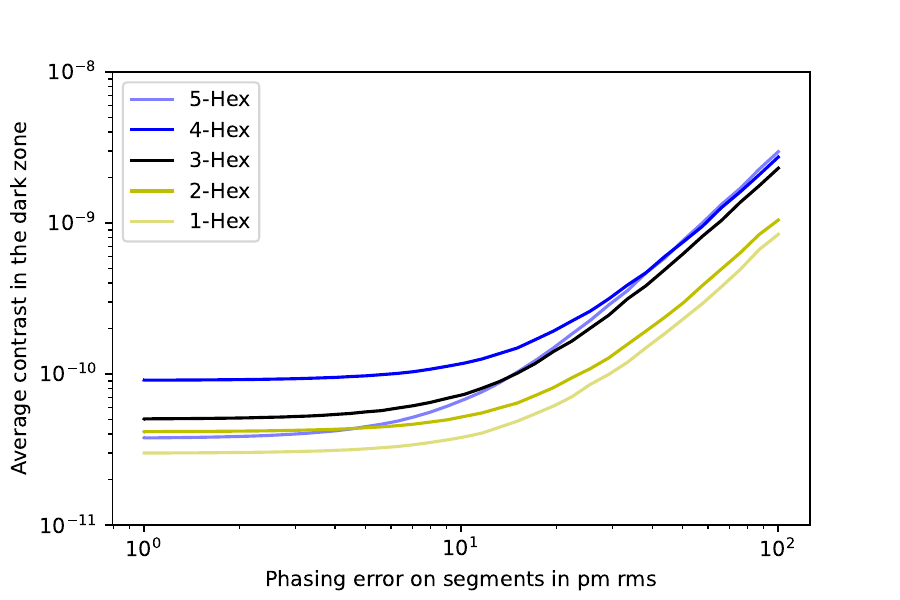}\\
   \includegraphics[width=8.5cm]{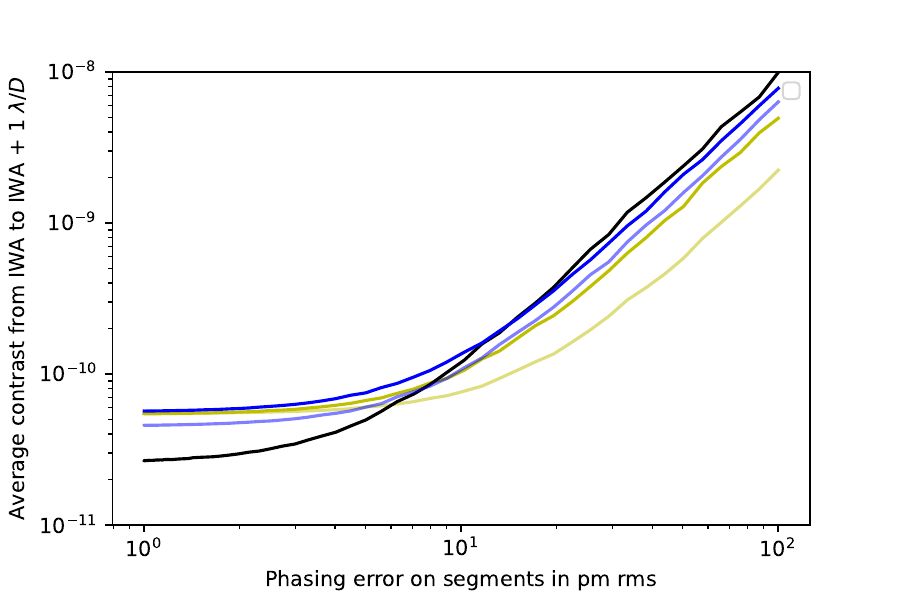}
   \end{tabular}
   \end{center}
   \caption[Fig] 
   { \label{fig:SEC4_Figure2_HockeyCross} Comparison of the error budgets of the five designs with respect to segment-level piston phasing errors: (top) average contrast in the dark region as a function of the wavefront error amplitude, (bottom) average contrast close to the IWA as a function of the wavefront error amplitude.}
   \end{figure}

As in the previous section, this effect is quantified using random segment-level piston errors with varying amplitudes, propagated through end-to-end simulations of the five coronagraphic designs. The simulations were performed using end-to-end simulations, with $100$ amplitude steps from $1$ to $100$ pm rms, and for each amplitude step, $150$ random segment phasing errors were generated and propagated through the system. Figure~\ref{fig:SEC4_Figure2_HockeyCross} shows the outcome performance and robustness of the five designs. Over the entire dark zone, for a contrast of $10^{-10}$, the constraints are relaxed by a factor of $1.9$ between 5-Hex and 1-Hex; and at small angular separations ($3.4-4.4 \lambda/D$), the constraints are released by a factor of $2$ between 5-Hex and 1-Hex for a contrast of $10^{-10}$.

\subsubsection{Robustness to segment-level tip–tilt}

Similarly to the piston case, the low-order envelopes associated with segment-level tip errors, shown in Fig.~\ref{fig:SEC4_Figure_DSP}, indicate that only the 1-Hex configuration does not have its peak within the dark region, implying an increased robustness to segment-level tip-tilt errors. In particular, at small angular separations, the envelopes of the 2-Hex to 5-Hex configurations are close to their maximum values.

\begin{figure}
   \begin{center}
   \begin{tabular}{c}
   \includegraphics[width=8.5cm]{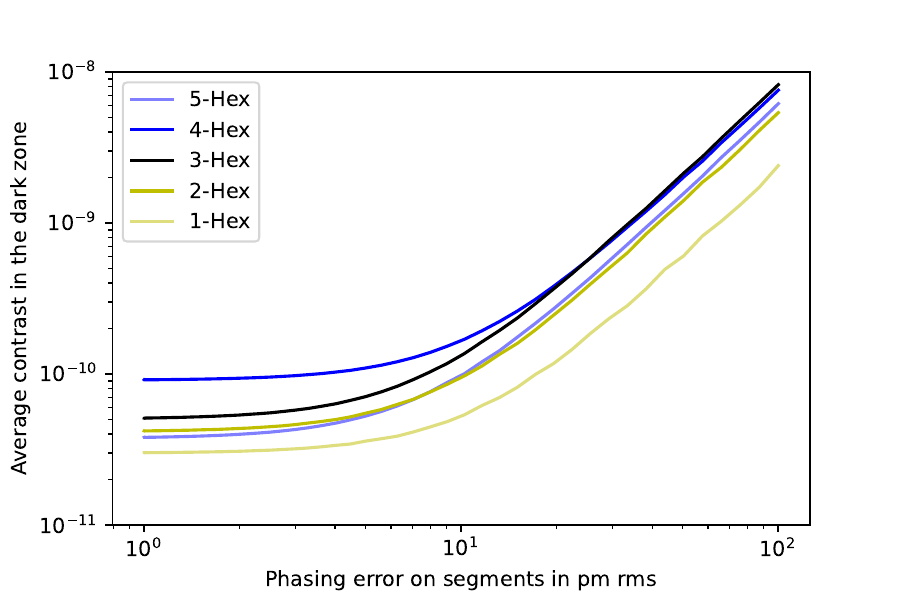}\\
   \includegraphics[width=8.5cm]{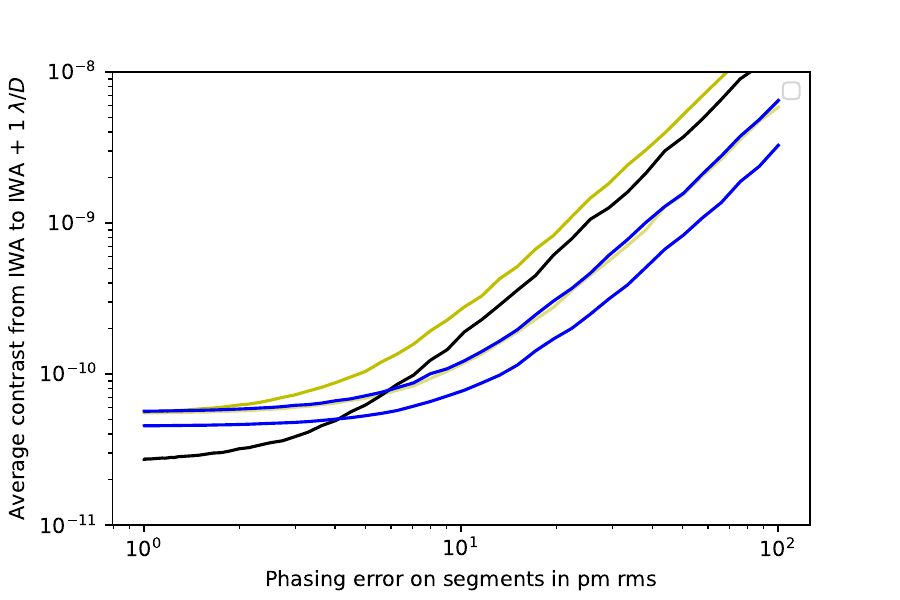}
   \end{tabular}
   \end{center}
   \caption[Fig] 
   { \label{fig:SEC4_Figure3_HockeyCross_TipTilt} Comparison of the error budgets of the five designs with respect to segment-level tip-tilt phasing errors: (top) average contrast in the dark region as a function of the wavefront error amplitude, (bottom) average contrast close to the IWA as a function of the wavefront error amplitude.}
   \end{figure}

Figure~\ref{fig:SEC4_Figure3_HockeyCross_TipTilt} shows the error budgets of the five designs as a function of increasing segment-level tip-tilt aberrations. Similarly to the previous section, it requires $100$ amplitude steps times $150$ random segment phasing errors end-to-end simulations, with amplitude steps from $1$ to $100$ pm rms. Over the entire dark zone, for a contrast of $10^{-10}$, the constraints are relaxed by a factor of $\sim1.5$ between 5-Hex and 1-Hex; and at small angular separations ($3.4-4.4 \lambda/D$), we cannot observe any tendency in terms of robustness.

As a conclusion, this section aimed to explore the segment-density parameter space while keeping the dark zone fixed, in contrast to Sect.~\ref{s:Impact of the focal-plane mask radius and system cutoff frequency}. The outcomes remain similar: a passive approach to moderately improving robustness consists of optimizing the segmentation scheme to keep the number of segments as small as possible, and in particular below the IWA expressed in $\lambda/D$.

\subsection{Phasing error sensing with a low-order wavefront sensor}

\begin{figure*}
   \begin{center}
   \includegraphics[width=16cm]{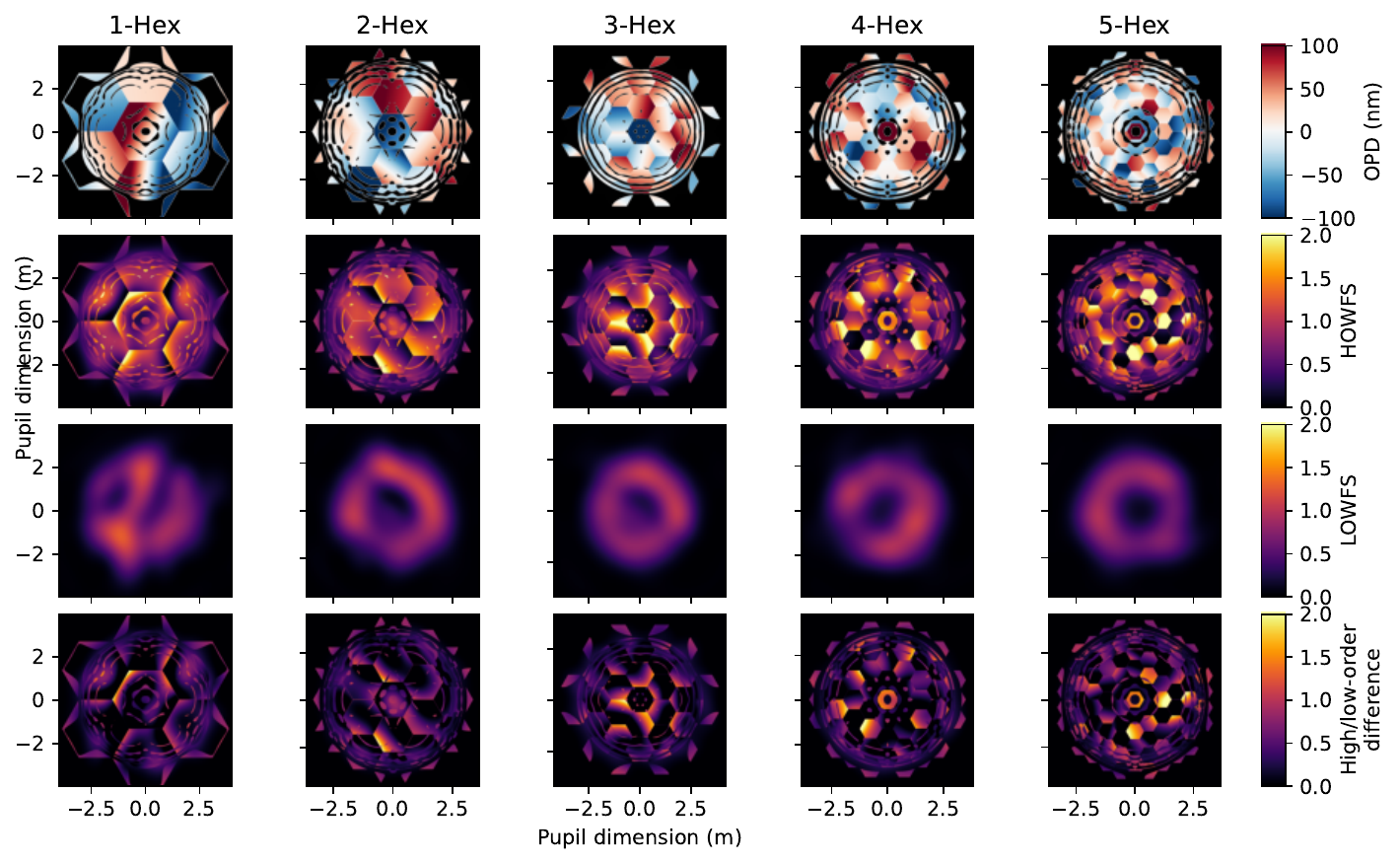}\
   \end{center}
   \caption[Fig] 
   { \label{fig:SEC4_Figure8_intensities} For the $5$ APLC cases, (row 1) input wavefront error (here $50$nm rms of segment-level piston-tip-tilt. These values are chosen artificially high for the ease of the visualization of the difference between the difference cases) and intensities of (row 2) their reconstructions through a high-order ZWFS (no filtering), (row 3) low-order ZWFS reconstructions, i.e. filtered by the FPM before reaching the ZWFS, (row 4) the difference of row 2 minus row 3.}
   \end{figure*}

For a specific phasing error, Figure~\ref{fig:SEC4_Figure8_intensities} illustrates the impact of segment density on the low-order ZWFS wavefront reconstruction, from 1-Hex to 5-Hex configurations, for a fixed FPM radius of $3.5\lambda/D$. The first row shows the input error phase map, with amplitudes of around $50$nm rms. The second and third rows respectively display the intensity maps reconstructed by an ideal high-order sensor (HOWFS, without FPM filtering) and by the low-order ZWFS (LOWFS), while the fourth row shows their absolute difference. As segment density increases (from 1-Hex to 5-Hex), the phasing error map contains increasingly high spatial frequencies, which have an impact outside of the FPM range.

\begin{figure}
   \begin{center}
   \includegraphics[width=8.5cm]{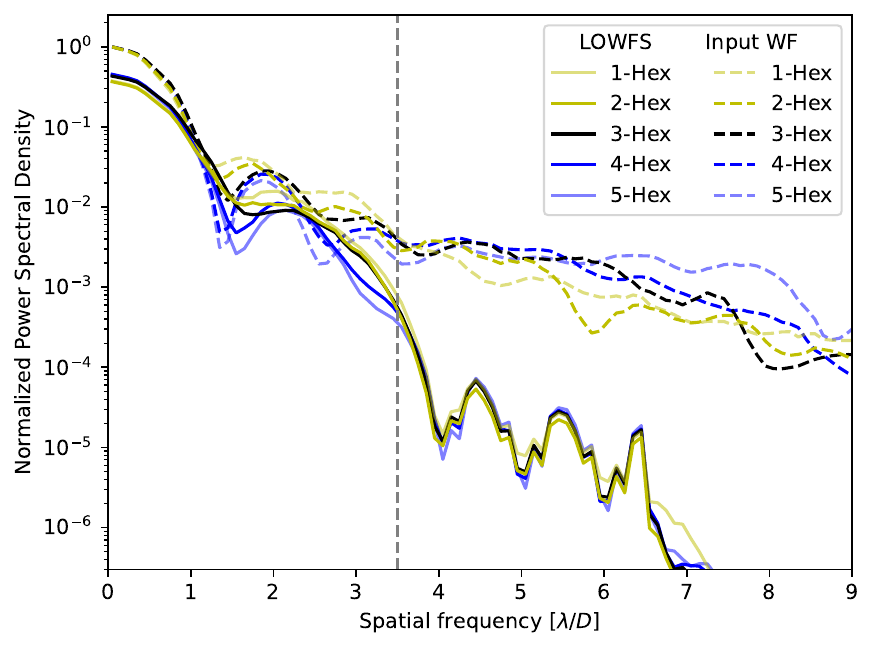}\
   \end{center}
   \caption[Fig] 
   { \label{fig:SEC4_Figure8_psds} Power spectral densities of the reconstructed aberrations for different segmentation types, compared to the ones of the input aberrated electric field.}
   \end{figure}

This spatial filtering effect is further quantified in Figure~\ref{fig:SEC4_Figure8_psds}, which shows the normalized PSD of the LOWFS-reconstructed wavefront electric field (solid lines) and of the input wavefront electric field (dashed lines) for each configuration. Let's note first that the input wavefront error PSDs are modulated by the pupil PSD, itself modulated by the segment envelope: as illustrated in Figure \ref{fig:Figure1_DSP}), only the core of the 1-Hex envelope is fully (in the piston case) to mostly (in the tip-tilt case) reflected by the FPM, which therefore captures the majority of the phasing error power. For all segment densities, the reconstructed PSDs accurately quite follows (what appears as a light scaling mismatch between the input electric field PSD and the reconstructed ones comes from the appearing pupil diameter being slightly reduced by the apodization of the external segments) the input PSDs at spatial frequencies below the FPM cutoff (marked by the vertical dashed line), before diverging above this threshold. Critically, as segment density increases from 1-Hex to 5-Hex, a larger fraction of the input wavefront power resides at spatial frequencies above the FPM cutoff, meaning that an increasingly significant portion of the phasing error remains unsensed by the LOWFS.

\begin{figure}
   \begin{center}
   \includegraphics[width=8.5cm]{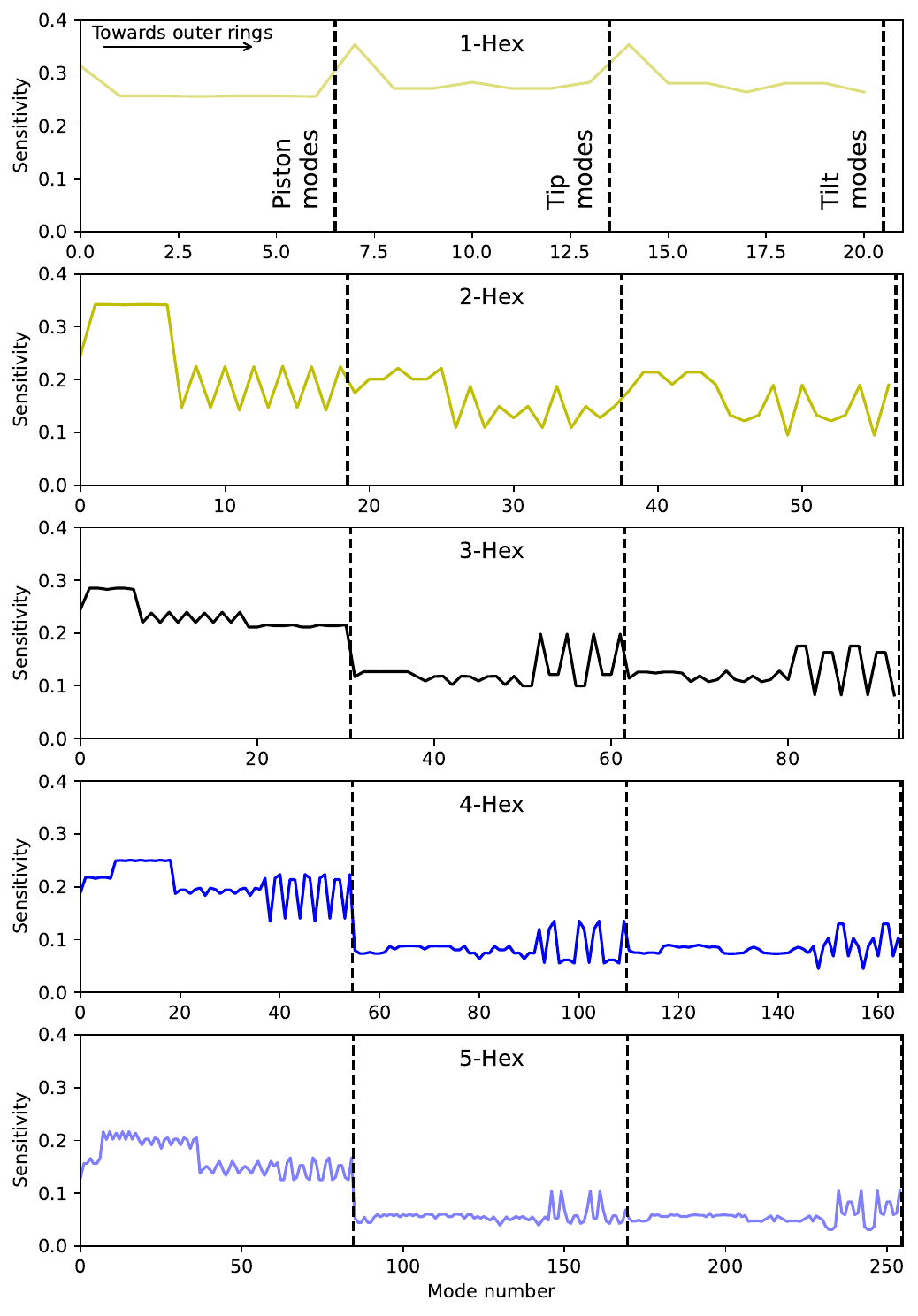}\
   \end{center}
   \caption[Fig] 
   { \label{fig:SEC4_Figure8_sensitivities} Sensitivity of the low-order ZWFS to segment piston, tip, and tilt modes as a function of mode number, for the five segmentation types. Within each mode family, the mode index increases from the innermost to the outermost segment ring.}
   \end{figure}

This behavior is reflected in the modal sensitivity analysis shown in Figure~\ref{fig:SEC4_Figure8_sensitivities}, where the LOWFS sensitivity is plotted as a function of mode number for segment-level piston, tip, and tilt, with mode indices increasing from inner to outer ring segments within each group. As in the previous section, several trends emerge consistently. First, for a given configuration, sensitivity varies from inner to outer ring modes: in all configurations, outer segments are more strongly apodized, as well as the central segment in most configurations. Second, piston modes exhibit higher sensitivity than tip and tilt modes across all configurations. Third, and most strikingly, the sensitivity over all modes decreases from 1-Hex to 5-Hex, by a factor of $\sim3$. Taken together, these results confirm that, for a fixed FPM radius, the LOWFS reconstruction performance degrades with increasing segment density. 

\section{Conclusions}
\label{s:Conclusions}

The study developed in this paper highlights how the combination of mirror segmentation, coronagraph architecture, and LOWFS implementation can be optimized to enhance robustness to segment phasing errors (piston, tip, and tilt), to achieve the extreme contrasts required for imaging close-in exoplanets.

In summary, segment-level piston, tip, and tilt errors produce low-order envelopes in the coronagraphic PSF, whose angular extent scales with the segment size relative to the primary mirror. The FPM acts as a spatial high-pass filter towards to science arm and its diameter plays a role in reducing the impact of these errors on the science contrast. Increasing the FPM radius from $3.5$ to $6.5\lambda/D$ relaxes piston phasing constraints by a factor of up to $3.5$ to maintain the contrast near the IWA, and tip-tilt constraints by a factor of up to $4.0$. Reducing the number of segments from $85$ (5-Hex configuration) to $7$ (1-Hex configuration) further relaxes piston constraints by a factor of $2$ to maintain contrast near the IWA, and tip-tilt constraints by a factor of $\sim 1.5$ over the full dark zone. It also affects the aberration signal reflected by the FPM and available to the LOWFS: the WFS more efficiently reconstructs segment-level errors when the mask radius is sufficiently large relative to the segment size. In our application case, increasing the FPM radius from $3.5$ to $6.5\lambda/D$ improves on average the sensitivity to segment modes by a factor of $\sim 2$, and decreasing the number of segments improves it by a factor of $\sim 3$. Overall, these results imply that passive design optimization can complement LOWFS correction, particularly at very small IWAs, helping to relax segment alignment requirements while maintaining the desired high contrast.

An interesting avenue for future development is multi-stage coronagraph architectures: by apodizing each segment to reduce its low-order envelope before combining it with an APLC, segment phasing constraints could be further relaxed, as suggested in \cite{Leboulleux2022}. However, the main limitation of the RAP approach is that it cannot access IWAs smaller than the segment diffraction limit, which further supports the case for a low-segment-density primary mirror.

It could also be interesting to explore the benefits of focal-plane wavefront sensing directly in the science image, such as pair-wise probing, while considering alternative coronagraph types (e.g., vortex or PIAACMC), also relevant to future missions like HWO. 
Overall, this system optimization involves a trade-off between two regimes. A small FPM or a high segment density primary mirror (more than $2 \times N$ segments across the pupil diameter, where $N$ is the coronagraph IWA in units of $\lambda/D$) would benefit focal-plane WFS approaches such as pair-wise probing, as segment phasing errors produce detectable signatures in the science image; yet the impact of such phasing errors on coronagraphic performance remains severe. On the other hand, a large FPM or a low segment density primary mirror (fewer than $2 \times N$ segments across its diameter) offers a dual advantage: phasing aberrations become detectable by a low-order ZWFS, while their impact on the science camera is intrinsically reduced, making this regime more favorable for both passive robustness and low-order WFS.

This study and this approach are directly relevant for telescopes with low segment density mirrors, where achieving the target performance (both in terms of IWA and contrast) imposes drastic segment phasing requirements. At the forefront of such missions, HWO aims for contrasts of $10^{-10}$ at very small angular separations (typically below $3 \lambda/D$ in the visible/NIR). This approach is particularly critical for a potential near-ultraviolet coronagraph, where the shorter wavelengths translate into tighter physical segment alignment tolerances for a given wavefront error budget. In this context, the relaxation offered by an optimized segmentation scheme and a properly sized FPM becomes especially valuable: by combining these two levers, the physical phasing tolerances in the NUV can be brought back to levels comparable to those required in the visible range, with the additional benefit of operating at a larger IWA in $\lambda/D$ units, i.e., a more favorable coronagraphic regime in terms of throughput and contrast, but also sensitivity to aberrations and wavefront sensing efficiency.

\begin{acknowledgements}
This study was initiated during L.P.’s visit to IPAG. The authors thank Université Grenoble-Alpes for funding this visit. L.L. also acknowledges support from the Action Spécifique Haute Résolution Angulaire (ASHRA), which funded a research stay at MPIA to finalize this work. \\
The aplc\_optimization package was created in support of the Segmented Coronagraph Design and Analysis (SCDA) study, funded by NASA's Exoplanet Exploration Program (ExEP). We wish to thank Space Telescope Science Institute collaborators, in particular, the SCDA team and Emiel Por, Kathryn St. Laurent, Remi Soummer, Mamadou N'Diaye, Remi Flamary, Bryony Nickson, Kelsey Glazer, James Noss, and Marshall Perrin.
\end{acknowledgements}

\bibliographystyle{aa}
\bibliography{bib}

\begin{appendix}
\section{Covariance matrices of the low-order ZWFS for segment-level piston, tip, and tilt errors}

\begin{figure}[ht!]
   \begin{center}
   \includegraphics[width=7cm]{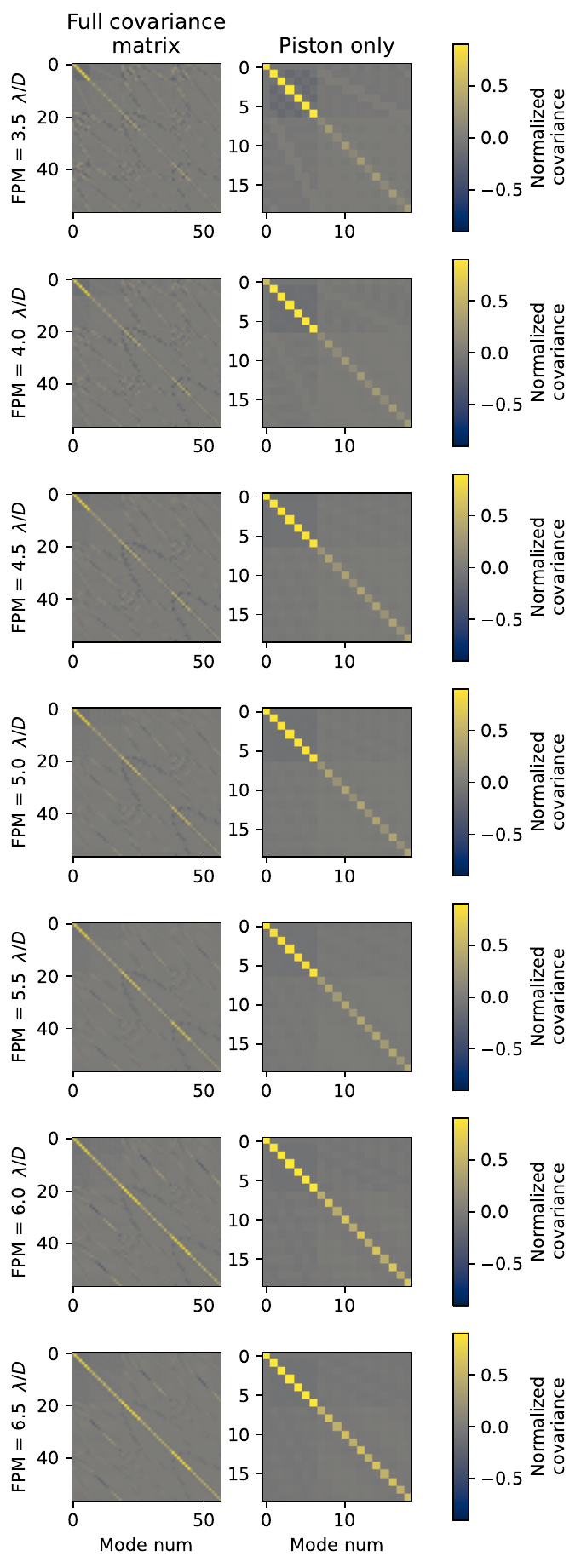}\
   \end{center}
   \caption[Fig] 
   { \label{fig:ANNEXE_covariance_fpm} Full covariance matrices (left column) and piston-only (right column) cropped from the full matrix of the low-order ZWFS reconstructor for segment-level piston, tip, and tilt modes, shown for the 2-Hex aperture and FPM radii ranging from $3.5$ to $6.5\lambda/D$. The matrices are normalized by their maximum in absolute value.}
   \end{figure}

\begin{figure}[ht]
   \begin{center}
   \includegraphics[width=7.5cm]{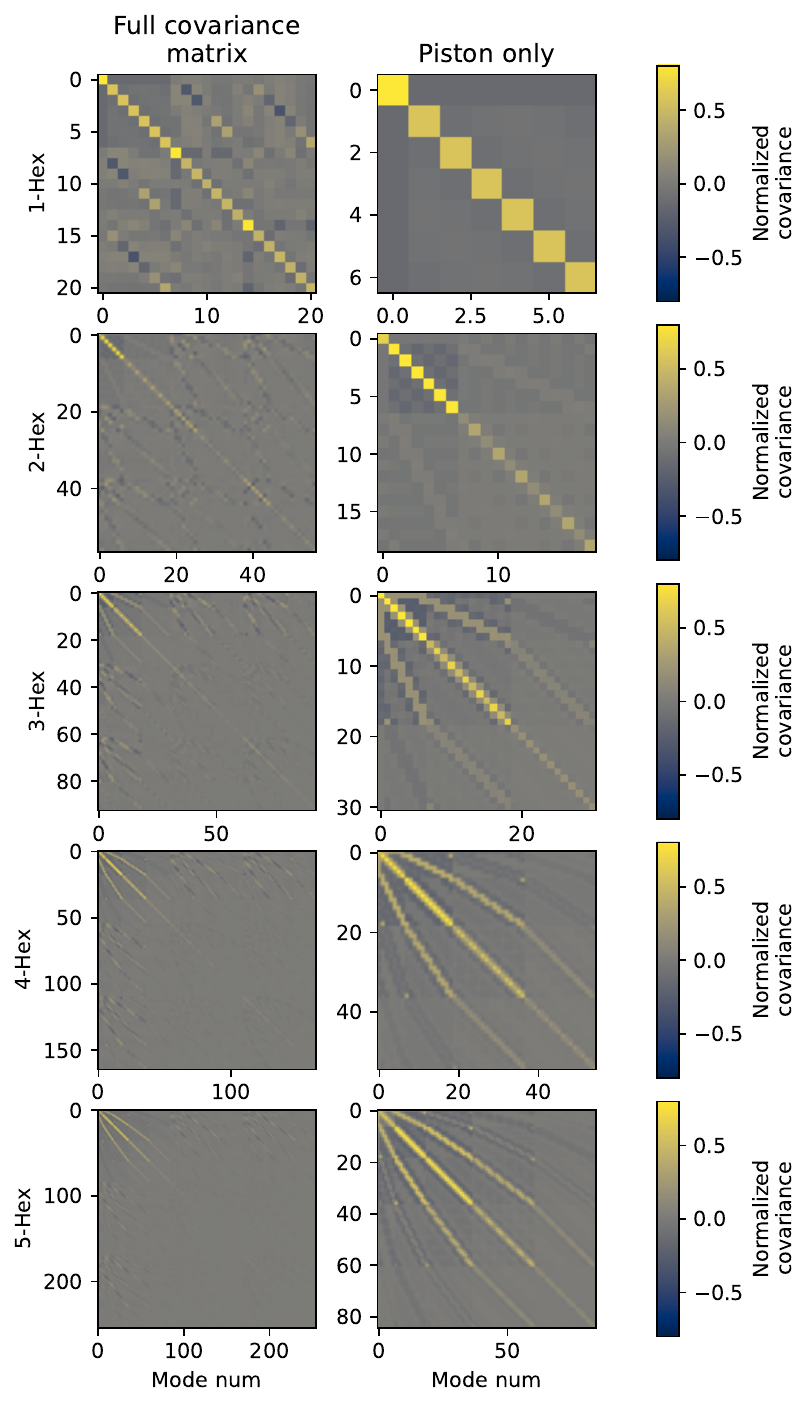}\
   \end{center}
   \caption[Fig] 
   { \label{fig:ANNEXE_covariance_segmentation} Full covariance matrices (left column) and piston-only (right column) cropped from the full matrix of the low-order ZWFS reconstructor for segment-level piston, tip, and tilt modes, shown for apertures ranging from 1-Hex to 5-Hex segmentation configurations. The matrices are normalized by their maximum in absolute value.}
   \end{figure}

Figures~\ref{fig:ANNEXE_covariance_fpm} and \ref{fig:ANNEXE_covariance_segmentation} show the full covariance matrices of the low-order ZWFS response to segment-level piston, tip, and tilt errors, respectively as a function of FPM radius (for the 2-Hex segmentation) and as a function of segmentation scheme (for a fixed FPM radius). The modes are ordered as in the main text: piston, tip, and tilt grouped by segment ring, with mode indices increasing from inner to outer rings. The diagonal terms correspond to the individual mode sensitivities already discussed in Figures~\ref{fig:Figure8_sensitivities} and \ref{fig:SEC4_Figure8_sensitivities}. Here we focus on the off-diagonal structure, which encodes cross-talk between modes, i.e., the degree to which the ZWFS response to one mode contaminates the reconstruction of another.

In Figure~\ref{fig:ANNEXE_covariance_fpm}, the full covariance matrices (left column) are strongly diagonal-dominated across all FPM radii, indicating that the ZWFS largely preserves mode orthogonality. However, residual off-diagonal terms are visible, with values up to $25-29\%$ of the maximum value (for all FPM radii), reflecting the spatial correlations introduced by the FPM low-order filtering. 

In the piston-only submatrices (right column), these off-diagonal terms are higher for small FPM radii, which can lead to mode confusion in the ZWFS reconstruction. As the FPM radius increases, this cross-talk is progressively reduced ($3\%$ to $0.07\%$ for a FPM radius from $3.5$ to $6.5\lambda/D$), further motivating the use of a larger FPM for robust segment phasing with a LOWFS, in the case of piston modes. Once again, this does not reflect the other segment modes.


This mode confusion is even more illustrated in Figure~\ref{fig:ANNEXE_covariance_segmentation}, where the covariance matrices are shown for increasing segmentation density (1-Hex to 5-Hex) at fixed FPM radius ($3.5\lambda/D$). For the 1-Hex case, the small number of segments results in a compact, well-conditioned matrix with limited off-diagonal contamination, and the piston-only submatrix (right column) is nearly block-diagonal. As the segmentation density increases, the off-diagonal structure becomes increasingly complex: cross-talk between modes of adjacent rings grows, and the piston-only submatrix develops extended off-diagonal features for the higher-density cases (3-Hex, 4-Hex, 5-Hex). From 1-Hex to 5-Hex apertures, the cross-talk covariances are of the same order of magnitude over all modes ($35-55\%$), but evolved from $14\%$ to $55\%$ for the piston modes only. This progressive degradation of mode orthogonality with increasing segment density is consistent with the FPM spatial filtering argument: for a fixed FPM radius, higher-density segmentations push more phasing modes above the FPM cutoff frequency, making them increasingly degenerate from the LOWFS perspective, and thus harder to disentangle in the wavefront reconstruction.


\end{appendix}

\end{document}